\documentclass[aps,prd,
nofootinbib,
preprintnumbers,
twocolumn,
showpacs, 
floatfix
]{revtex4-1}
\usepackage{amsmath}
\usepackage{amssymb}
\usepackage{epsfig}
\usepackage{graphicx}
\usepackage{multirow}
\usepackage{color}
\usepackage[normal]{subfigure}
\usepackage{rotating}
\usepackage{hyperref}

\newcommand{\capdef}{}
\newcommand{\mycaption}[2][\capdef]{\renewcommand{\capdef}{#2}
\caption[#1]{{\footnotesize #2}}}

\newcommand{\ewt}{\end{widetext}}
\newcommand{\be}{\begin{equation}}
\newcommand{\ee}{\end{equation}}
\newcommand{\bdm}{\begin{displaymath}}
\newcommand{\edm}{\end{displaymath}}
\newcommand{\bea}{\begin{eqnarray}}
\newcommand{\eea}{\end{eqnarray}}
\newcommand{\nn}{\nonumber}

\def\eq#1{{Eq.~(\ref{#1})}}
\def\eqs#1#2{{Eqs.~(\ref{#1})--(\ref{#2})}}

\def\Table#1{{Table~\ref{#1}}}

\def\app#1{{Appendix~\ref{#1}}}

\def\vev#1{\left\langle #1 \right\rangle}

\def\Tr{\mbox{Tr}\,}

\begin{document}
\preprint{IFIC/12-07, TTP12-004, SFB/CPP-12-06}
\title{Seesaw Scale in the Minimal Renormalizable $SO(10)$ Grand Unification}
\author{Stefano Bertolini}\email{bertolin@sissa.it}
\affiliation{INFN, Sezione di Trieste, SISSA,
via Bonomea 265, 34136 Trieste, Italy}
\author{Luca Di Luzio}\email{diluzio@particle.uni-karlsruhe.de}
\affiliation{Institut f\"{u}r Theoretische Teilchenphysik,
Karlsruhe Institute of Technology (KIT), D-76128 Karlsruhe, Germany}
\author{Michal Malinsk\'{y}}\email{malinsky@ific.uv.es}
\affiliation{AHEP Group, Instituto de F\'{\i}sica Corpuscular -- C.S.I.C./Universitat de Val\`encia, Edificio de Institutos de Paterna, Apartado 22085, E 46071 Val\`encia, Spain}
\begin{abstract}
Simple $SO(10)$ Higgs models with the adjoint representation triggering the grand-unified symmetry breaking, discarded a long ago due to inherent tree-level tachyonic instabilities in the physically interesting scenarios, have been recently brought back to life by quantum effects. In this work we focus on the variant with $45_{H}\oplus 126_{H}$ in the Higgs sector and show that there are several regions in the parameter space of this model that can support stable unifying configurations with the $B-L$ breaking scale as high as $10^{14}$~GeV, well above the previous generic estimates based on the minimal survival hypothesis. This admits for a renormalizable implementation of the canonical seesaw and makes the simplest potentially realistic scenario of this kind a good candidate for a minimal $SO(10)$ grand unification. 
Last, but not least, this setting is likely to be extensively testable at future large-volume facilities such as Hyper-Kamiokande. 
\end{abstract}
\maketitle

\section{Introduction}
For the last about thirty years, the simplest non-supersymmetric $SO(10)$ gauge models with $45_H \oplus 16_H$ or $45_H \oplus 126_H$ in the Higgs sector have been widely considered uninteresting for any realistic unified model building. This was namely due to the tachyonic instabilities in their tree-level spectra popping up in all settings  compatible with the basic gauge unification constraints \cite{Buccella:1980qb,Yasue:1980fy,Yasue:1980qj,Anastaze:1983zk,Babu:1984mz} which, in non-SUSY settings, generically favour intermediate-energy thresholds. 
However, as it was shown recently in \cite{Bertolini:2009es,Bertolini:2010ng}, such instabilities are just artefacts of the tree-level approximation. Hence, technically, quantum effects bring this class of models back from oblivion. 

On the other hand, dedicated renormalization group studies such as~\cite{Chang:1984qr,Deshpande:1992au,Deshpande:1992em,Bertolini:2009qj} reveal that a successful unification in this class of models typically requires the $B-L$ breaking scale below $10^{12}$ GeV for the $45_H \oplus 16_H$ variant and below $10^{10}$ GeV in the $45_H \oplus 126_H$ case. Such values, however, are disfavoured by the neutrino oscillation and cosmology data: 
{\it i)} In the former case, $\vev{16_H}$ breaks the $B-L$ symmetry by one unit and, thus, the seesaw requires a pair of  $\vev{16_H}$ insertions. This can be minimally implemented at the renormalizable level by e.g.~a variant of the Witten's radiative mechanism \cite{Witten:1979nr,Bajc:2004hr,Bajc:2005aq} or,  giving up renormalizability, by a $d=5$ operator. In either case the ``effective'' $\Delta(B-L) = 2$ seesaw scale is further suppressed with respect to the $B-L$ breaking scale and the  light neutrino masses are typically overshoot by many orders of magnitude. Moreover, the non-renormalizable nature of the seesaw in the $d=5$ case hinders the general predictivity of this model.
{\it ii)} With $126_H$ at play, the $B-L$ symmetry is broken by two units so the right-handed neutrinos receive their masses at the tree level via the renormalizable $16_{F}16_{F}126^{*}_{H}$ Yukawa interaction~\cite{Babu:1992ia,Bajc:2005zf}. The upper limit on $\vev{126_H}$ quoted above then again pushes the absolute scale of the light neutrino masses much above the current limits. 

Though unpleasant, this, however, does not constitute a fundamental blow to the minimal non-SUSY $SO(10)$ as an extensive multi-parameter fine-tuning in the seesaw formula can still bring the light neutrino masses down to the desired sub-eV domain. In this respect, the situation is very different from that of the minimal supersymmetric (SUSY) $SO(10)$ grand unified theory (GUT)~\cite{Clark:1982ai,Aulakh:1982sw,Aulakh:2003kg,Fukuyama:2004pb,Fukuyama:2004xs,Goh:2003sy,Goh:2003hf,Bajc:2004xe,Aulakh:2005bd} where the neutrino masses are typically undershot; indeed, the rigidity of the Higgs potential in minimal SUSY Higgs models {\em enforces} a population pseudo-Goldstone bosons well below the GUT scale~($M_{G}$)~\cite{Bajc:2004xe} whenever the $SO(10)\to \text{SM}$ breaking is not essentially one-step~\cite{Aulakh:2005mw,Bertolini:2006pe}, 
hence disturbing the nearly ideal unification within the Minimal Supersymmetric Standard Model (MSSM); for further information see, e.g., \cite{Bajc:2008dc} and references therein. 

In the same spirit, one should keep in mind that the key upper bounds on the $B-L$ scale identified in~\cite{Chang:1984qr,Deshpande:1992au,Deshpande:1992em,Bertolini:2009qj} are derived under the strong assumption of the minimal survival hypothesis~\cite{delAguila:1980at}, i.e., that a minimal set of needed intermediate thresholds cluster exactly at the relevant symmetry breaking scale. This, of course, does not need to be the case in general and as little as a single unexpected multiplet in the bulk can open a room for $B-L$ scales much above the naive expectation, thus rendering the gauge coupling unification compatible with the neutrino data for a reasonable price. In this respect, the non-SUSY models with higher-dimensional Higgs representations (such as $45_H \oplus 126_H$) featuring a number of free parameters in the Higgs potential\footnote{Here the non-SUSY nature of the model is central - the SM-vacuum manifold of the minimal SUSY GUT, as complicated as it naively looks, is in reality very simple; indeed, it is parametrized by a single complex parameter~\cite{Bajc:2004xe}.} provide a lot of room for such a serendipity. Moreover, given the renormalizable nature of the seesaw in the $45_H \oplus 126_H$ case, the Yukawa sector of this kind of models is strongly constrained, which further opens the door for their near future testability. 

In this study we focus on the possible role of accidental thresholds in the desert of the minimal $SO(10)$ GUTs based on the $45_H \oplus 126_H$ Higgs sector. In particular, we calculate the tree-level spectrum of the minimal Higgs model and the leading universal radiative correction to the relevant Higgs masses and ask ourselves {\it i)}~whether states with accidentally small masses can pop up in some regions of the parametric space without destabilising the scalar potential and {\it ii)}~whether the corresponding threshold effects can lift the seesaw scale to the desired ballpark of $10^{13 \div 14}$~GeV.

The work is organized as follows: In Section~\ref{45126Higgsmodel} we define the $45_H \oplus 126_H$ $SO(10)$ Higgs model of interest and calculate its tree-level spectrum\footnote{Though there exist detailed studies of the vacuum of the $54_H \oplus 126_H$ Higgs model, cf.~\cite{Buccella:1984ft,Buccella:1986hn}, 
a similar analysis for the setting with $45_H$ instead of $54_H$, to our best knowledge, has never been done. 
}, which reveals the expected tachyonic instabilities except for the phenomenologically questionable $SU(5)$-like descents. In analogy to the canonical example elaborated on in \cite{Bertolini:2009es} we argue that radiative corrections alleviate the issue and that stable and potentially realistic SM vacua are accessible. To exemplify that, we calculate the leading $SO(10)$-invariant radiative correction as a minimal scalar-spectrum regulator. In Section~\ref{sect:unification} we study the possible effects of various multiplets -- if they happen to live in the ``GUT desert'' -- on the actual location of the $B-L$ scale. We identify two specific simple and consistent settings in which all  current phenomenological constraints from the proton decay searches and big-bang nucleosynthesis are compatible with the latest limits on the absolute neutrino mass scale. A simple numerical scan over the parametric space reveals extended domains supporting these solutions. Remarkably enough, in both cases the extra threshold is pinned to a relatively narrow mass window which, in turn, yields a rather specific prediction for the position of the GUT scale and, hence, the $d=6$ proton decay rate, well within the reach of the future large volume facilities such as Hyper-Kamiokande (Hyper-K)~\cite{Abe:2011ts}.

With all this at hand, in Section~\ref{sect:newminimalSO10} we make a case for a new potentially realistic minimal renormalizable $SO(10)$ GUT based on the $45_H \oplus 126_H \oplus 10_H$ Higgs sector. We comment in brief on the prospects and strategies of a future more detailed scrutiny of the scheme, paying particular attention to the Yukawa sector fits and the ultimate calculation of the proton decay branching ratios in the fully consistent settings. Then we conclude. Technical aspects of the Higgs and gauge-boson spectrum calculation are deferred to a set of Appendices. 

\section{The 45-126 Higgs Model}
\label{45126Higgsmodel}

\subsection{The tree-level scalar potential}
\label{treescalpot}

The most general renormalizable scalar potential that can be written with $45_H$ and $126_H$ at hand reads
\be
\label{scalpotgen}
V = V_{45} + V_{126} + V_{\rm mix} \, ,
\ee
where
\bea
\label{V45}
V_{45} &=& - \frac{\mu^2}{2} (\phi \phi)_0 + \frac{a_0}{4} (\phi \phi)_0 (\phi \phi)_0 + \frac{a_2}{4} (\phi \phi)_2 (\phi \phi)_2 \, , \\ \nn \\
\label{V126}
V_{126} &=&  - \frac{\nu^2}{5!} (\Sigma \Sigma^*)_0\\
& +& \frac{\lambda_0}{(5!)^2} (\Sigma \Sigma^*)_0 (\Sigma \Sigma^*)_0 
+ \frac{\lambda_2}{(4!)^2} (\Sigma \Sigma^*)_2 (\Sigma \Sigma^*)_2\nn\\
& + & \frac{\lambda_4}{(3!)^2(2!)^2} (\Sigma \Sigma^*)_4 (\Sigma \Sigma^*)_4
+ \frac{\lambda'_{4}}{(3!)^2} (\Sigma \Sigma^*)_{4'} (\Sigma \Sigma^*)_{4'} \nn \\ \nn \\
&+& \frac{\eta_2}{(4!)^2} (\Sigma \Sigma)_2 (\Sigma \Sigma)_2
+ \frac{\eta_2^*}{(4!)^2} (\Sigma^* \Sigma^*)_2 (\Sigma^* \Sigma^*)_2 \, , \nn \\ \nn \\
\label{V45126}
V_{\rm mix} &=& \frac{i \tau}{4!} (\phi)_2 (\Sigma \Sigma^*)_2 
+ \frac{\alpha}{2 \cdot 5!} (\phi \phi)_0 (\Sigma \Sigma^*)_0 \\
&+& \frac{\beta_4}{4 \cdot 3!} (\phi \phi)_4 (\Sigma \Sigma^*)_4
+ \frac{\beta'_{4}}{3!} (\phi \phi)_{4'} (\Sigma \Sigma^*)_{4'} \nn \\ \nn \\ 
&+& \frac{\gamma_2}{4!} (\phi \phi)_2 (\Sigma \Sigma)_2
+ \frac{\gamma_2^*}{4!} (\phi \phi)_2 (\Sigma^* \Sigma^*)_2 \,\nn .
\eea
Here we have used the symbols $\phi$ and $\Sigma$ for the components of $45_H$ and $126_H$, respectively. The detailed breakdown of all the contractions (with the subscripts denoting the number of open indices in the relevant brackets) is given 
in~\app{app:details}. 
Notice that all the couplings are real but $\eta_2$ and $\gamma_2$.

\subsection{The symmetry breaking patterns}
\label{sec:breakingpatterns}

\label{sec:SMsinglet}

There are in general three SM singlets in the reducible $45_H\oplus126_H$ representation of $SO(10)$. 
Using $BL \equiv (B-L)/2$
and labelling the field components with respect to the
$3_{c}\, 2_{L}\, 2_{R}\, 1_{BL}$ (i.e.,  $SU(3)_c \otimes SU(2)_L \otimes SU(2)_R \otimes U(1)_{BL}$) algebra, the SM singlets
reside in the $(1,1,1,0)$ and $(1,1,3,0)$ sub-multiplets of $45_{H}$
and in the $(1,1,3,+1)$ component of $126_{H}$.
In what follows we shall denote
\be
\label{vevs}
\vev{(1,1,1,0)}\equiv \omega_{BL},\, \vev{(1,1,3,0)}\equiv \omega_{R} ,\, \vev{(1,1,3,+1)}\equiv \sigma,
\ee
where $\omega_{BL,R}$ are real and $\sigma$ can be made real by
a phase redefinition of the $126_H$. 
Different VEV configurations
trigger the spontaneous breakdown of the $SO(10)$ symmetry into several qualitatively distinct subgroups. Namely, for $\sigma= 0$ one finds (in an obvious notation)
\begin{align}
\label{vacua}
&\omega_{R}= 0,\, \omega_{BL}\neq 0\; : & 3_c\, 2_L\, 2_R\, 1_{BL}\,, \nn \\[0.5ex]
&\omega_{R}\neq 0,\, \omega_{BL}= 0\; : & 4_{C} 2_L 1_R\,, \nn \\[0.5ex]
&\omega_{R}\neq 0,\, \omega_{BL}\neq 0\; : & 3_c\, 2_L\, 1_R\, 1_{BL}\,,\\[0.5ex] 
&\omega_{R}=-\omega_{BL}\neq 0\; : & \mbox{flipped}\, 5'\, 1_{Z'}\,, \nn \\[0.5ex]
&\omega_{R}=\omega_{BL}\neq 0\; :  & \mbox{standard}\, 5\, 1_{Z}\,, \nn
\end{align}
with  $5\, 1_{Z}$ and $5'\, 1_{Z'}$ standing for the two inequivalent
embeddings of the SM hypercharge operator $Y$ into $SU(5) \otimes U(1)\subset SO(10)$ usually called the ``standard'' and the ``flipped'' $SU(5)$ scenarios~\cite{DeRujula:1980qc,Barr:1981qv}, respectively. 
In the standard case, 
$
Y=T^{(3)}_R+T_{BL}
$
belongs to the $SU(5)$ algebra and the orthogonal Cartan generator
$Z$ is given by
$
Z =-4T^{(3)}_R+6T_{BL}
$.
In the flipped ($5'1_{Z'}$) case, the right-handed isospin
assignment of quarks and leptons 
is turned over so that the flipped hypercharge generator reads
$Y'=-T^{(3)}_R+T_{BL}$.
Accordingly, the additional $U(1)_{Z'}$ generator reads
$Z' =4T^{(3)}_R+6T_{BL}$ (for further details see ,e.g.,~Ref.~\cite{Bertolini:2009es}).

For $\sigma \neq 0$ all the intermediate gauge symmetries (\ref{vacua}) are spontaneously broken down to the SM group, with the exception of the last case which maintains the $SU(5)$ subgroup unbroken and, hence, will not be considered here.
The decomposition of the $45_H$ and $126_H$ representations with respect to the all relevant intermediate symmetries~(\ref{vacua}) is 
detailed in Tables \ref{tab:45decomp} and \ref{tab:126decomp}.

\renewcommand{\arraystretch}{1.25}
\begin{table*}
\begin{tabular}{ccccc|cc}
\hline \hline
 $4_C\,2_L\,2_R $
& $4_C\,2_L\,1_R $
& $3_c\,2_L\,2_R\,1_{BL} $
& $3_c\,2_L\,1_R\,1_{BL} $
& $3_c\,2_L\,1_Y $
& $5\,1_Z$ 
& $5'\,1_{Z'}$ 
\\
\hline
 $\left({ 1,1,3} \right)$
& $\left({ 1,1},+1 \right)$
& $\left({ 1,1,3},0 \right)$
& $\left({ 1,1},+1,0 \right)$
& $\left({ 1,1},+1 \right)$
& $\left({ 10},-4 \right)$
& $\left({ \overline{10}},+4 \right)$
\\
\null
& $\left({ 1,1},0 \right)$
&
& $\left({ 1,1},0,0 \right)$
& $\left({ 1,1},0 \right)$
& $\left({ 1},0 \right)$
& $\left({ 1},0 \right)$
\\
\null
& $\left({ 1,1},-1 \right)$
&
& $\left({ 1,1},-1,0 \right)$
& $\left({ 1,1},-1 \right)$
& $\left({ \overline{10}},+4 \right)$
& $\left({ 10},-4 \right)$
\\
$\left({ 1,3,1} \right)$
& $\left({ 1,3},0 \right)$
& $\left({ 1,3,1},0 \right)$
& $\left({ 1,3},0,0 \right)$
& $\left({ 1,3},0 \right)$
& $\left({ 24},0 \right)$
& $\left({ 24},0 \right)$
\\
$\left({ 6,2,2} \right)$
& $\left({ 6,2},+\frac{1}{2} \right)$
& $\left({ 3,2,2},-\frac{1}{3} \right)$
& $\left({ 3,2},+\frac{1}{2},-\frac{1}{3} \right)$
& $\left({ 3,2},\frac{1}{6} \right)$
& $\left({ 10},-4 \right)$
& $\left({ 24},0 \right)$
\\
\null
& $\left({ 6,2},-\frac{1}{2} \right)$
&
& $\left({ 3,2},-\frac{1}{2},-\frac{1}{3} \right)$
& $\left({ 3,2},-\frac{5}{6} \right)$
& $\left({ 24},0 \right)$
& $\left({ 10},-4 \right)$
\\
\null
&
& $\left({ \overline{3},2,2},+\frac{1}{3} \right)$
& $\left({ \overline{3},2},+\frac{1}{2},+\frac{1}{3} \right)$
& $\left({ \overline{3},2},+\frac{5}{6} \right)$
& $\left({ 24},0 \right)$
& $\left({ \overline{10}},+4 \right)$
\\
\null
&
&
& $\left({ \overline{3},2},-\frac{1}{2},+\frac{1}{3} \right)$
& $\left({ \overline{3},2},-\frac{1}{6} \right)$
& $\left({ \overline{10}},+4 \right)$
& $\left({ 24},0 \right)$
\\
$\left({ 15,1,1} \right)$
& $\left({ 15,1},0 \right)$
& $\left({ 1,1,1},0 \right)$
& $\left({ 1,1},0,0 \right)$
& $\left({ 1,1},0 \right)$
& $\left({ 24},0 \right)$
& $\left({ 24},0 \right)$
\\
\null
&
& $\left({ 3,1,1},+\frac{2}{3} \right)$
& $\left({ 3,1},0,+\frac{2}{3} \right)$
& $\left({ 3,1},+\frac{2}{3} \right)$
& $\left({ \overline{10}},+4 \right)$
& $\left({ \overline{10}},+4 \right)$
\\
\null
&
& $\left({ \overline{3},1,1},-\frac{2}{3} \right)$
& $\left({ \overline{3},1},0,-\frac{2}{3} \right)$
& $\left({ \overline{3},1},-\frac{2}{3} \right)$
& $\left({ 10},-4 \right)$
& $\left({ 10},-4 \right)$
\\
\null
&
& $\left({ 8,1,1},0 \right)$
& $\left({ 8,1},0,0 \right)$
& $\left({ 8,1},0 \right)$
& $\left({ 24},0 \right)$
& $\left({ 24},0 \right)$
\\
\hline \hline
\end{tabular}
\mycaption{Decomposition of the adjoint representation $45$ with respect to the various $SO(10)$ subgroups. 
The definitions and normalization of the abelian charges are given in the text.}
\label{tab:45decomp}
\end{table*}

\renewcommand{\arraystretch}{1.25}
\begin{table*}
\begin{tabular}{ccccc|cc}
\hline \hline
 $4_C\,2_L\,2_R $
& $4_C\,2_L\,1_R $
& $3_c\,2_L\,2_R\,1_{BL} $
& $3_c\,2_L\,1_R\,1_{BL} $
& $3_c\,2_L\,1_Y $
& $5\, 1_Z$ 
& $5'\, 1_{Z'}$
\\
\hline
$\left({ 6,1,1} \right)$
& $\left({ 6,1,0} \right)$
& $\left({ \overline{3},1,1},+\frac{1}{3} \right)$
& $\left({ \overline{3},1,0},+\frac{1}{3} \right)$
& $\left({ \overline{3},1},+\frac{1}{3} \right)$
& $\left({ \overline{5}},+2\right)$
& $\left({ \overline{5}},+2\right)$
\\
\null
& 
& $\left({ 3,1,1},-\frac{1}{3} \right)$
& $\left({ 3,1,0},-\frac{1}{3} \right)$
& $\left({ 3,1},-\frac{1}{3} \right)$
& $\left({ 45},-2\right)$
& $\left({ 45},-2\right)$
\\
\null
$\left({ 10,3,1} \right)$
& $\left({ 10,3,0} \right)$ 
& $\left({ 1,3,1},-1 \right)$
& $\left({ 1,3,0},-1 \right)$
& $\left({ 1,3},-1 \right)$
& $\left({ \overline{15}},-6\right)$
& $\left({ \overline{15}},-6\right)$
\\
\null
& 
& $\left({ 3,3,1},-\tfrac{1}{3} \right)$
& $\left({ 3,3,0},-\tfrac{1}{3} \right)$
& $\left({ 3,3},-\tfrac{1}{3} \right)$
& $\left({ 45},-2\right)$
& $\left({ 45},-2\right)$
\\
\null
& 
& $\left({ 6,3,1},+\tfrac{1}{3} \right)$
& $\left({ 6,3,0},+\tfrac{1}{3} \right)$
& $\left({ 6,3},+\tfrac{1}{3} \right)$
& $\left({ \overline{50}},+2\right)$
& $\left({ \overline{50}},+2\right)$
\\
\null
$\left({ \overline{10},1,3} \right)$
& $\left({ \overline{10},1,-1} \right)$ 
& $\left({ 1,1,3},+1 \right)$
& $\left({ 1,1,-1},+1 \right)$
& $\left({ 1,1},0 \right)$
& $\left({ 1},+10\right)$
& $\left({ \overline{50}},+2\right)$
\\
\null
& $\left({ \overline{10},1,0} \right)$ 
& 
& $\left({ 1,1,0},+1 \right)$
& $\left({ 1,1},+1 \right)$
& $\left({ 10},+6\right)$
& $\left({ 10},+6\right)$
\\
\null
& $\left({ \overline{10},1,+1} \right)$ 
& 
& $\left({ 1,1,+1},+1 \right)$
& $\left({ 1,1},+2 \right)$
& $\left({ \overline{50}},+2\right)$
& $\left({ 1},+10\right)$
\\
\null
& 
& $\left({ \overline{3},1,3},+\tfrac{1}{3} \right)$
& $\left({ \overline{3},1,-1},+\tfrac{1}{3} \right)$
& $\left({ \overline{3},1},-\tfrac{2}{3} \right)$
& $\left({ 10},+6\right)$
& $\left({ 45},-2\right)$
\\
\null
& 
& 
& $\left({ \overline{3},1,0},+\tfrac{1}{3} \right)$
& $\left({ \overline{3},1},+\tfrac{1}{3} \right)$
& $\left({ \overline{50}},+2\right)$
& $\left({ \overline{50}},+2\right)$
\\
\null
& 
& 
& $\left({ \overline{3},1,+1},+\tfrac{1}{3} \right)$
& $\left({ \overline{3},1},+\tfrac{4}{3} \right)$
& $\left({ 45},-2\right)$
& $\left({ 10},+6\right)$
\\
\null
& 
& $\left({ \overline{6},1,3},-\tfrac{1}{3} \right)$
& $\left({ \overline{6},1,-1},-\tfrac{1}{3} \right)$
& $\left({ \overline{6},1},-\tfrac{4}{3} \right)$
& $\left({ \overline{50}},+2\right)$
& $\left({ \overline{15}},-6\right)$
\\
\null
& 
& 
& $\left({ \overline{6},1,0},-\tfrac{1}{3} \right)$
& $\left({ \overline{6},1},-\tfrac{1}{3} \right)$
& $\left({ 45},-2\right)$
& $\left({ 45},-2\right)$
\\
\null
& 
& 
& $\left({ \overline{6},1,+1},-\tfrac{1}{3} \right)$
& $\left({ \overline{6},1},+\tfrac{2}{3} \right)$
& $\left({ \overline{15}},-6\right)$
& $\left({ \overline{50}},+2\right)$
\\
\null
$\left({ 15,2,2} \right)$
& $\left({ 15,2,-\tfrac{1}{2}} \right)$ 
& $\left({ 1,2,2},0 \right)$
& $\left({ 1,2,-\tfrac{1}{2}},0 \right)$
& $\left({ 1,2},-\tfrac{1}{2} \right)$
& $\left({ \overline{5}},+2\right)$
& $\left({ 45},-2\right)$
\\
\null
& $\left({ 15,2,+\tfrac{1}{2}} \right)$ 
&
& $\left({ 1,2,+\tfrac{1}{2}},0 \right)$
& $\left({ 1,2},+\tfrac{1}{2} \right)$
& $\left({ 45},-2\right)$
& $\left({ \overline{5}},+2\right)$
\\
\null
&
& $\left({ \overline{3},2,2},-\tfrac{2}{3} \right)$
& $\left({ \overline{3},2,-\tfrac{1}{2}},-\tfrac{2}{3} \right)$
& $\left({ \overline{3},2},-\tfrac{7}{6} \right)$
& $\left({ 45},-2\right)$
& $\left({ \overline{15}},-6\right)$
\\
\null
&
& 
& $\left({ \overline{3},2,+\tfrac{1}{2}},-\tfrac{2}{3} \right)$
& $\left({ \overline{3},2},-\tfrac{1}{6} \right)$
& $\left({ \overline{15}},-6\right)$
& $\left({ 45},-2\right)$
\\
\null
&
& $\left({ 3,2,2},+\tfrac{2}{3} \right)$
& $\left({ 3,2,+\tfrac{1}{2}},+\tfrac{2}{3} \right)$
& $\left({ 3,2},+\tfrac{7}{6} \right)$
& $\left({ \overline{50}},+2\right)$
& $\left({ 10},+6\right)$
\\
\null
&
& 
& $\left({ 3,2,-\tfrac{1}{2}},+\tfrac{2}{3} \right)$
& $\left({ 3,2},+\tfrac{1}{6} \right)$
& $\left({ 10},+6\right)$
& $\left({ \overline{50}},+2\right)$
\\
\null
& 
& $\left({ 8,2,2},0 \right)$
& $\left({ 8,2,-\tfrac{1}{2}},0 \right)$
& $\left({ 8,2},-\tfrac{1}{2} \right)$
& $\left({ \overline{50}},+2\right)$
& $\left({ 45},-2\right)$
\\
\null
& 
&
& $\left({ 8,2,+\tfrac{1}{2}},0 \right)$
& $\left({ 8,2},+\tfrac{1}{2} \right)$
& $\left({ 45},-2\right)$
& $\left({ \overline{50}},+2\right)$
\\
\hline \hline
\end{tabular}
\mycaption{Same as in Table~\ref{tab:45decomp} for the $126$ representation.}
\label{tab:126decomp}
\end{table*}

\subsection{The tree-level scalar spectrum}\label{treescalarspectrum}
Adopting the convention in which the mass term in the Lagrangian is written as $\tfrac{1}{2}\psi^T M^2 \psi$, where $\psi=(\phi, \Sigma^\ast, \Sigma)$ is a 297-dimensional vector, 
the scalar spectrum is obtained readily by evaluating the relevant functional scalar mass matrix of the schematic form
\be
M^2(\phi, \Sigma^*, \Sigma)=\left(
\begin{array}{lll}
V_{\phi\phi} & V_{\phi\Sigma^\ast} & V_{\phi\Sigma} \\
V_{\Sigma^\ast\phi} & V_{\Sigma^\ast\Sigma^\ast} & V_{\Sigma^\ast\Sigma} \\
V_{\Sigma\phi} & V_{\Sigma\Sigma^\ast} & V_{\Sigma\Sigma}
\end{array}
\right)
\label{M2matrix}
\ee
 on the SM vacuum.
The subscripts here denote the derivatives of the scalar potential with respect to a specific set of fields.
Subsequently, this matrix is brought to a block-diagonal form by a subsequent unitary transformation 
into the SM basis.

The complete tree-level spectrum is given in Appendix~\ref{app:tree_spectrum}. There are several features that can be seen readily: {\it i)} as anticipated in~\cite{Bertolini:2009es} there is again a pair of pseudo-Goldstone bosons (cf.~also comments in Appendix~\ref{globalsymmetries})  entertaining very simple mass formulae: 
\bea
\label{treemass130}
M^2 (1,3,0) &=& - 2 a_2 (\omega_{BL} - \omega_R) (\omega_{BL} + 2 \omega_R) \, , \\
\label{treemass810}
M^2 (8,1,0) &=& - 2 a_2 (\omega_R - \omega_{BL}) (\omega_R + 2 \omega_{BL}) \,.
\eea
These multiplets develop tachyonic masses whenever $\omega_{BL}/\omega_{R}$ is outside the $[-2,-\tfrac{1}{2}]$ interval.
Hence, as such, the tree-level Higgs spectrum is clearly unable to support the physically interesting breaking patterns with either 
$\omega_{BL}\ll \omega_{R}$ or $\omega_{R}\ll \omega_{BL}$, thus avoiding the intermediate flipped $SU(5)' \otimes U(1)_{Z'}$ stage. 
$SU(5)$ intermediate stages, $\omega_{BL}, \omega_{R} \ll \sigma$, are disfavoured as well\footnote{Recently, there were several attempts to reconcile the simplest non-SUSY $SU(5)$ scenarios with the gauge unification by means of intermediate-scale thresholds, see, e.g., \cite{Dorsner:2005fq,Dorsner:2005ii,Dorsner:2006dj,Bajc:2006ia,Bajc:2007zf,Dorsner:2006fx,Dorsner:2007fy} and references therein. Though possible in principle, we shall not consider the $SU(5)$ option here because there is virtually no room for such an intermediate stage below the $SO(10)$-breaking scale in the model of our interest. Moreover, these settings generically rely on several multiplets pushed into the desert.}.

{{\it ii)} In this respect, it is worth looking at formulae (\ref{treemass130}) and (\ref{treemass810}) in more detail. For instance, as in the $45_H \oplus 16_H$ case~\cite{Bertolini:2009es} there are no contributions there from the $B-L$ --breaking VEV $\sigma$ although the number of  available contractions of the type $(\phi^{2})(\Sigma\Sigma^{(*)})$ is larger here. This can be understood as follows: regardless of how the indices of the $\Sigma\Sigma^{(*)}$ bilinears are contracted, the resulting tensor never breaks the $SU(5)$ symmetry. Since both $\Sigma$'s couple to both $\phi$'s in the same manner one can always view the contraction with the pair of adjoints ($\phi$'s) as a quadratic covariant-derivative-like term for the fields with the SM gluon and $SU(2)_{L}$-gauge quantum numbers. These fields, however,  remain massless at the $SU(5)$ level.

However, as shown in~\cite{Bertolini:2009es}, this is no longer the case at the quantum level where all the tree-level forbidden couplings do indeed enter the relevant mass formulae and thus open the room for the physically interesting settings with $\omega_{BL}$ very different from $\omega_{R}$. 
\subsection{Leading one-loop corrections\label{sect:loopcorrections}}
Unlike in the $45_H \oplus 16_H$ case the full-fledged effective potential (EP) calculation of the one-loop scalar spectrum in the $45_H \oplus 126_H$ model is very difficult due to the enormous complexity of the contractions involving $126_H$ so we shall not attempt it here. However, the radiative corrections are really important only for the pseudo-Goldstone bosons associated to  accidental global symmetries, c.f.~\cite{Bertolini:2009es} and Appendix~\ref{globalsymmetries}; thus, one can get a good grip on the one-loop spectrum even without the full EP analysis. Moreover, some of the results obtained for the $45_H \oplus 16_H$ setting in~\cite{Bertolini:2009es} can be readily adopted to the current case; in particular, the one-loop gauge-induced corrections to the masses of the scalars residing solely in $45_{H}$  (such as $(8,1,0)$ and $(1,3,0)$) are identical to those obtained  in \cite{Bertolini:2009es}, cf.~formulae~(D1)-(D2) therein. 
This, however, is not the case for the contribution of scalars which span over the components of $126_H$. Similarly, the one-loop scalar-induced contributions to the tree level scalar masses should be calculated from scratch. Needless to say, this is a formidable task if it is to be performed in full generality. 

Thus, in what follows, we shall focus only on the most universal scalar one-loop correction, namely, the leading non-logarithmic $SO(10)$-invariant $\tau^{2}$-proportional term which, as we argue, can be fully accounted for by a simple diagrammatic calculation. Since it yields a positive correction to all the scalar masses, it should already be enough to regularize the salient tachyonic instabilities of the tree-level scalar spectrum and, perhaps, open new regions in the parametric space where stable unifying configurations with  phenomenologically favourable intermediate scales could be supported. Moreover, since also the other leading non-logarithmic corrections (i.e., those coming from the gauge and the remaining scalar loops) are typically positive, including just the $SO(10)$ invariant piece can be viewed as a minimalistic attempt to stabilize the tachyons.
In view of this, a detailed calculation of all one-loop corrections to the scalar spectrum in this framework is not even necessary and will be left to a dedicated future study. 

Since the leading scalar-loop induced non-logarithmic corrections in the scalar sector come from tadpoles~\cite{Aoki:1982ed}, it is easy to see that the only source of a $\tau^{2}$-proportional non-log term is associated to the renormalization of the stationarity conditions. Diagrammatically, it corresponds to a special cluster of one-loop graphs contributing to the one-point function of $45_{H}$ of the kind
\be
\parbox{2.4cm}{\includegraphics[width=2.4cm]{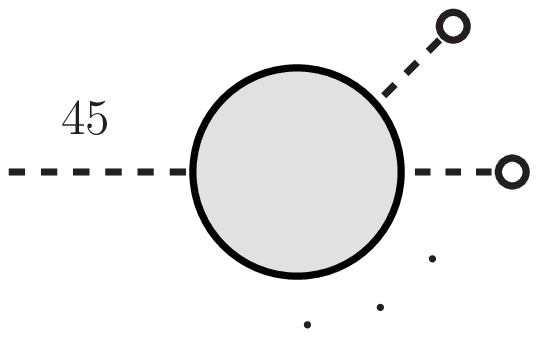}}\ni
\parbox{2.6cm}{\includegraphics[width=2.6cm]{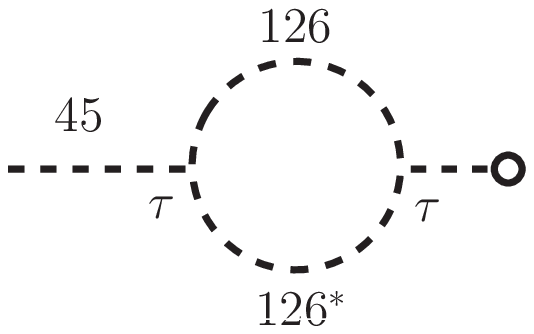}}+
\parbox{2.4cm}{\includegraphics[width=2.4cm]{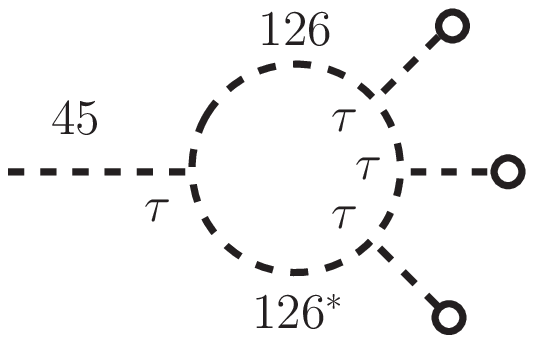}}+\ldots
\ee
Given the $SO(10)$ structure of the relevant $\tau$-vertex, namely, 
\be\label{tausquaredcorrection}
V_{45-126}\ni \frac{i\tau}{4!}\phi_{ij} \Sigma_{klmni}\Sigma^{*}_{klmnj}
\ee
the universal mass shift due to this class of graphs reads\footnote{Let us just mention that the same technique applied to the $45_H \oplus 16_H$ case yields the uniform mass shift of $\tau^{2}/4\pi^{2}$ which is indeed consistent with the results of the effective potential calculation~\cite{Bertolini:2009es}.}
\be\label{leadingoneloopcorrection}
\Delta M^{2}_{\text{1-loop-}\tau^{2}}=\frac{35\tau^{2}}{32\pi^{2}}+\text{logs}\,,
\ee 
where the symbol ``logs'' denotes all the logarithmic corrections that are minimized at  the GUT scale. 
\section{Unification in the 45-126  model}\label{sect:unification}
With this information at hand, in this section we can finally address the question of our main interest, namely, whether accidentally light scalar multiplets in the SM desert could possibly open the door to a consistent gauge unification with a $B-L$ scale well above the unpleasant upper limit of about $10^{10}$ GeV obtained in~\cite{Bertolini:2009qj} under the assumption of minimal survival. 

Since the scalar masses are expressed as functions of the microscopic parameters entering the scalar potential~(\ref{scalpotgen}), pushing a specific multiplet into the desert amounts to imposing an extra algebraic constraint on the parameter space of the model, i.e., it cuts out a region close to the relevant zero-mass hyper-surface. The rest of the spectrum then must be evaluated around this hyper-surface which, however, brings in a high level of non-linearity. Thus, in what follows, we shall mainly stick to numerical methods to simulate the heavy scalar and vector-boson spectra in order to single out the regions of the parametric space that can support viable gauge unification patterns. 
\subsection{Consistency}\label{sect:consistency}
Besides gauge unification, there are other basic aspects of an overall consistency of  potentially realistic settings that will be of our concern here, namely, the stability of the physical vacuum (i.e., the absence of tachyons) and the position of the unification scale\footnote{The GUT scale is conventionally defined as the mass scale of the gauge bosons associated to the breakdown of $SO(10)$ to either $3_{c}2_{L}2_{R}1_{BL}$ or $4_{C}2_{L}1_{R}$ intermediate symmetries, i.e., those transforming for instance as $(3,2,-\tfrac{5}{6})\oplus (\overline{3},2,+\tfrac{5}{6})$ under the SM gauge group.} which governs the $d=6$ proton decay. Moreover, with potentially very light coloured states in the desert, $d>6$ proton decay as well as possible BBN issues should be also considered.   

\subsubsection{Vacuum stability}\label{sect:vacuumstability}
As stated before, from now on we shall stick to the ``minimally regularized'' form of the scalar spectrum, i.e., we shall use the tree-level formulae of Appendix~\ref{scalspectSM} augmented with the leading $SO(10)$-invariant non-logarithmic one-loop correction (\ref{leadingoneloopcorrection}). 

For each physical point in the parametric space, the mass-squares of all propagating scalars should be positive. This, as we shall see, is indeed a very restrictive constraint which already disqualifies some of the potentially interesting multiplets, see Sect.~\ref{sect:suitablemultiplets}.  
It is perhaps worth mentioning that with one such a stable vacuum at hand one can generate a continuum of other stable vacua by rescaling all the dimensionful parameters entering the mass formulae by a common factor. This invariance will be later on used for a simple optimization of the one-loop unification patterns, cf.~Sect.~\ref{sect:technicalities}. Moreover, it is easy to understand that, as long as only the scalar mass-squares are concerned, further degeneracies in the parametric space of the model can be identified; among these perhaps the most prominent is the absence of the phase of $\gamma_{2}$ from the tree-level mass formulae and the irrelevance of the overall  sign of the mass parameters at play (i.e., all that matters are just relative signs). 

\subsubsection{\label{sect:protonlimits}Proton lifetime limits}
\paragraph{$d=6$ proton decay:}
We shall impose the latest (2011) Super-Kamiokande (SK) limit on the proton lifetime (for the $e^{+}\pi^{0}$ 
channel)~\cite{Nakamura:2010zzi}:
\be\label{limit:SK2011}
\tau(p\to e^{+}\pi^{0})_{\rm SK, 2011}> 8.2 \times 10^{33}\, {\rm years}\,,
\ee
and, whenever appropriate, comment on the changes in the results for a couple of assumed future sensitivity limits, namely those quoted in~\cite{Abe:2011ts} that 
Hyper-Kamiokande (HK) should reach by 2025 and 2040, respectively:
\bea
\label{limit:HK2025}\tau(p\to e^{+}\pi^{0})_{\rm HK, 2025}& > &  9 \times 10^{34}\, {\rm years}\,,\\
\label{limit:HK2040}\tau(p\to e^{+}\pi^{0})_{\rm HK, 2040} & > &  2 \times 10^{35}\, {\rm years}\,.
\eea
These translate to the following (raw) formula for the compatibility regions in the $M_{G}-\alpha_{G}^{-1}$ plane:
\be\label{protondecaylimits}
\left(\frac{\alpha_{G}^{-1}}{45}\right)10^{2(n_{G}-15)}>11.8,\; 39.0,\; 58.1,
\ee 
where $n_{G}\equiv \log_{10}(M_{G}/{\rm GeV})$ and the three values on the right-hand-side correspond to the three lifetime limits in \eqs{limit:SK2011}{limit:HK2040}, respectively.
In the relevant figures (cf. FIGs~\ref{FinetunedOmegaBL63p13}-\ref{FIGOmegaRsigma63p13} and FIGs~\ref{FinetunedOmegaR82p12}-\ref{FIGOmegaBLsigma82p12}), the regions of the parametric space where the three constraints (\ref{protondecaylimits}) are fulfilled will be, consecutively, denoted by light-gray, dark-gray and a black color.  

One should also check that lowering a specific multiplet into the GUT desert  does not bring any of the proton-dangerous coloured scalar triplets too much below some $10^{14}$~GeV; although the detailed structure of the scalar $d=6$ proton decay amplitude is typically suppressed by small Yukawa couplings, this is not always the case and a coloured triplet well below this limit can be dangerous. Since we do not consider the details of the Yukawa sector here, we shall adopt a conservative limit like the one quoted above. Remarkably enough, this constraint turns out to be rather weak and in a vast majority of the cases where (\ref{protondecaylimits}) are obeyed the scalar triplets are safe.   
\vskip 1mm
\paragraph{$d>6$ proton decay:\label{d7protondecay}}
Under the ``big desert'' hypothesis the $d=6$ proton decay operators 
conserve $B-L$ up to $M_W / M_G$ 
corrections~\cite{Weinberg:1979sa,Wilczek:1979hc}\footnote{In the $SO(10)$ models these operators are usually induced by the scalar triplets transforming as 
$(3,1,-\tfrac{1}{3}) \oplus (\overline{3},1,+\tfrac{1}{3})$ and the 
$(3,2,-\tfrac{5}{6}) \oplus (\overline{3},2,+\tfrac{5}{6}) \oplus 
(3,2,+\tfrac{1}{6}) \oplus (\overline{3},2,-\tfrac{1}{6})$ gauge bosons.}.
However this picture does not need to hold anymore if we consider new 
structures at intermediate scales well below $M_G$ 
and $d>6$ proton decaying operators (such as those conserving $B+L$ at the $d=7$ level, c.f.~\cite{Wilczek:1979et,Weldon:1980gi}) should be inspected. A ``canonical'' example here is the situation when the $(3,2,+\tfrac{1}{6})$ scalar approaches the weak scale; the relevant $B+L$ conserving proton decay amplitude\footnote{In the current $SO(10)$ model the relevant effective operator is traced back to the $126_H^4$  quartic coupling and the $16_F 16_F 126_H^*$ Yukawa 
interaction.} can then easily clash with the experimental limits~\cite{Dorsner:2005fq}.

\subsubsection{\label{sect:BBNconstraints}BBN and the lifetime of light coloured BSM multiplets}
Light colored thresholds can be also troublesome for the Big Bang Nucleosynthesis (BBN). 
This has to do with the requirement that any colored state other than the SM fields
must decay with a lifetime shorter than about $1$~second, in order to preserve the classical predictions of the light elements'
abundances~\cite{Nakamura:2010zzi}. From this perspective, renormalizable Yukawa couplings of such light scalars to the SM matter fields are welcome as the relevant decay widths are typically large enough to be safe.

\subsection{Running with extra thresholds in the desert}
For the sake of simplicity, in what follows we shall entirely stick to the case with a single extra SM sub-multiplet of $45_H \oplus 126_H$ in the desert. This not only lowers the number of fine-tunings to the minimum, but also admits for a systematic classification of the possible threshold effects. 
\subsubsection{Identifying the most suitable thresholds\label{sect:suitablemultiplets}}
{\it i)} The stability requirements of~Sect.~\ref{sect:vacuumstability} disfavour a light $(3,3,-\tfrac{1}{3})$ multiplet as there are no suitable stable vacua supporting this configuration even if the leading universal one-loop correction (\ref{tausquaredcorrection}) is taken into account\footnote{Strictly speaking, this is not entirely decisive as the other one-loop corrections we are not considering here may open more room for such a setting.}.

{\it ii)} There is a good reason to disfavour all multiplets whose effect on the hypercharge coupling evolution is much larger than the effect on the $SU(2)_{L}$ coupling: 
Recall that the upper limit on $B-L$ emerges from the need to delay the ``premature'' SM unification of the $U(1)_{Y}$ and $SU(2)_{L}$ couplings by lowering enough the $B-L$ scale. An extra state in the desert which would act against this rule of thumb would further strengthen the demands imposed on the $B-L$ scale, thus further lowering the relevant upper bound. 
On the other hand, such states are almost never brought down alone as the relevant fine-tuning lowers also the states occupying the same larger-symmetry multiplets to the respective symmetry breaking scale; however, such intermediate scales in the settings of our interest should not be far from~$M_{G}$ so the effects of these extra components are typically sub-leading.
Hence, multiplets like $(1,1,+1)$, $(1,1,+2)$, $(\overline{3},1,-\tfrac{2}{3})$, $(\overline{3},1,+\tfrac{4}{3})$, $(\overline{6},1,-\tfrac{4}{3})$, $(\overline{6},1,-\tfrac{1}{3})$, $(\overline{6},1,+\tfrac{2}{3})$ and $(3,2,+\tfrac{7}{6})$ are not fit for our purposes. On the same footing, the individual effect of an additional 
$(1,2,+\tfrac{1}{2})$ is too weak to make much difference even if it is pushed down to the electroweak scale.

{\it iii)} We discard also the $(1,3,-1)$ component of $126_{H}$ because it is the type-II seesaw triplet -- indeed, a very light triplet would require an extra fine-tuning of the effective $SU(2)_{L}$-triplet-doublet-doublet coupling otherwise the absolute neutrino mass scale would be overshot by many orders of magnitude. 

Thus, from now on we shall entirely focus on the possible threshold effects due to the remaining SM multiplets pushed into the GUT desert, namely, the  $(1,3,0)$ submultiplet of $45_{H}$, a pair of $(3,2,+\tfrac{1}{6})$ mixed multiplets, the $(6,3,+\tfrac{1}{3})$ of $126_{H}$ and the pair of $(8,2,+\tfrac{1}{2})$ in $126_{H}$.

\subsubsection{Technical details of the RGE analysis}
\label{sect:technicalities}
On the technical side, we shall always work in the effective SM picture where all the beyond-SM scalar and vector bosons are classified by their $3_{c}2_{L}1_{Y}$ quantum numbers; hence, conveniently, we will be always using the three SM ``effective'' couplings irrespective of the actual number of simple gauge factors that can be identified at any given energy scale. Needless to say, this is a mere convention provided that the matching to the full theory (especially at higher orders) is performed consistently. 

Given the tachyonic nature of the tree-level spectrum in the settings of our interest, a pure one-loop RGE analysis is meaningless; in principle, the simplest fully consistent approach would be, of course, a two-loop running based on a complete one-loop information about the scalar and gauge spectra. 

This, however, is extremely demanding in full generality because even the very analytic  minimization of the relevant one-loop effective potential in the $45_H \oplus 126_H$ case is virtually intractable (note that, in this respect, the qualitative difference between the $45_H \oplus 16_H$ and $45_H \oplus 126_H$ cases is paramount). 

Thus, we shall rather perform a qualitative one-loop RGE analysis based on the "minimally regularised" scalar spectrum, see~Sections~\ref{sect:loopcorrections} and \ref{sect:vacuumstability} which, however, should\footnote{This expectation is based on the simple fact (see for instance~\cite{Bertolini:2009es}) that the tachyons, which can be identified with pseudo-Goldstone modes of accidental global symmetries of the scalar potential restored in certain corners of the parametric space, are the only states whose mass-squares are really prone to radiative corrections.} account for all the salient features of the fully consistent picture. In other words, we work in the approximation in which the full one-loop approach to the gauge coupling evolution is refined by the key two-loop effects. 

Technically, the calculations are performed in three stages along the following lines: First, we randomly scan over the parametric space of the model assuming the desired multiplet to be close to the electroweak scale and calculate the scalar spectrum for each such a point. For those points for which the  vacuum turns out to be stable, we adjust the overall scale of the dimensionful parameters $\omega_{R}$, $\omega_{BL}$, $\sigma$ and $\tau$ and the position of the light threshold in such a way that a consistent unification is obtained. Note that, in many cases, this can be done even analytically -- at one loop, both such changes inflict essentially linear shifts in the values of the three gauge couplings evaluated above the highest threshold\footnote{Concerning the latter, i.e., the shift in the mass of the accidentally light multiplet, as long as it is well below the next-to-lightest and other heavier thresholds it can slide essentially freely without affecting the heavy part of the spectrum at all because all such configurations fall into a very small patch of the parametric space. This, however, does not need to be always the case and one should be more careful here, cf.~FIG.~\ref{Stability_63p13}.} so the optimization of the unification pattern amounts to a solution of a linear system. 
Finally, we check the full consistency of the resulting pattern with the proton decay and BBN limits specified in Section~\ref{sect:consistency} and see whether the threshold effects can lift the $B-L$ scale into the seesaw-favoured domain of $10^{13\div 14}$~GeV. 
\vskip 2mm 

Given that, one can identify the following main sources of uncertainties plaguing the precision of the derived electroweak-scale values of the gauge couplings:
{\it i)} Sticking to the one-loop beta-functions we commit an error of the size of a typical two-loop effect\footnote{More precisely, here we refer to the typical scale of the two-loop corrections in ``regular'' settings, i.e., those in which one does not encounter tachyonic instabilities and a one-loop analysis would be perfectly self-consistent.}. Assuming the usual size of such an uncertainty as observed, e.g., in~\cite{Bertolini:2009qj} one can expect a reduction of the tree-level prediction of $M_{G}$ by roughly a factor of two. {\it ii)} We do not re-input the derived values of the gauge couplings back into the gauge-boson mass-formulae and reiterate the code; for the sake of simplicity, we rather use a ``typical'' gauge coupling corresponding to $\alpha^{-1}_{G}$ around 40 for all heavy gauge boson masses. The error due to this is of the order of $ \log[g^2_{\rm true}/g^2_{\rm assumed}]/16\pi^2$ which, however, is entirely negligible. {\it iii)} We treat the fine-tuning in the $(1,2,+\tfrac{1}{2})$ sector in a simplified manner: since one eigenstate of the $(1,2,+\tfrac{1}{2})$ system is implicitly assumed to be fine-tuned to the electroweak scale to act as the usual SM Higgs boson only the heavier eigenstate of the doublet mass matrix should be included in the heavy-spectrum analysis. On the other hand, it does not make sense to perform such a fine-tuning with just $45_H \oplus 126_H$ at play as it would, artificially, bring in an extra constraint on the parametric space which would be, however, absent in any realistic model including e.g.~an extra $10_H$ in the Higgs sector. For the sake of this qualitative analysis, we decided to resolve this dichotomy by mimicking the effect of the (unidentified) heavy doublet by averaging\footnote{Note that this way one implicitly retains the desired slope ($b_{i}=-{37}/{3}$) for all three effective SM couplings above the scale of the heaviest component of the scalar spectrum.} over the effects of the two massive eigenstates of the doublet mass matrix~(\ref{matrix:12p12}). Note also that the effect of a possible extra $10_{H}$ in the full-fledged models (like those discussed later in Sect.~\ref{sect:newminimalSO10}) is expected to be small because the extra degrees of freedom would typically cluster around the GUT scale  and, amounting to a full irreducible $SO(10)$ representation, they would affect the GUT-scale position only marginally. Last, but not least, an extra $10_{H}$ at play does not contain a new candidate for a suitable low-scale threshold so, in this respect, the classification given in Sect.~\ref{sect:suitablemultiplets} is not affected.   

\subsection{Results}
Let us begin with a short comment on the first two options identified in Sect.~\ref{sect:suitablemultiplets}, namely, a light $(1,3,0)$ and/or a light component of the $(3,2,+\tfrac{1}{6})$ scalar pair. 
Although in both cases one can find regions of the parametric space supporting such light multiplets in the desert, the predicted position of the GUT scale is always at least an order of magnitude below the current Super-K limit (cf.~Sect.~\ref{sect:protonlimits}) and, hence, the $d=6$ proton decay constraints are always badly violated. Thus, none of these two cases turns out to be interesting.

However, as we argue below, in the remaining cases, namely, with either $(6,3,+\tfrac{1}{3})$ or $(8,2,+\tfrac{1}{2})$ in the desert, fully consistent solutions do exist. Moreover, in both these settings, {\it the upper limit on the $B-L$ scale is pushed up by several orders of magnitude with respect to the naive estimate based on the extended survival hypothesis}~\cite{Bertolini:2009qj}, thus opening a room for a natural implementation of the renormalizable seesaw in this class of $SO(10)$~GUTs. 
\subsubsection{Consistent setting 1: light $(6,3,+\tfrac{1}{3})$}
\paragraph{Stable vacua with a light $(6,3,+\tfrac{1}{3})$:}
Let us first quote the results of a dedicated numerical scan of the parametric space aiming at the identification of the stable vacua supporting an accidentally light $(6,3,+\tfrac{1}{3})$. Confining all the dimensionless couplings into the conservative $[-1,+1]$ range stable vacua are confined in the domain $\omega_{BL} >0$, $\beta_4'<0$, $\beta_4 > 0$, $a_{0}>-0.1$ and $|\gamma_2|<0.6$.
\paragraph{One-loop unification with a light $(6,3,+\tfrac{1}{3})$:}
Sample regions of the parametric space that support a consistent scalar spectrum and, at the same time, provide a viable gauge-coupling unification, are depicted in FIGs.~\ref{FinetunedOmegaBL63p13}, \ref{FIGGUTscale63p13} and  \ref{FIGOmegaRsigma63p13}.   
\begin{figure}[ht]
\includegraphics[width=7cm]{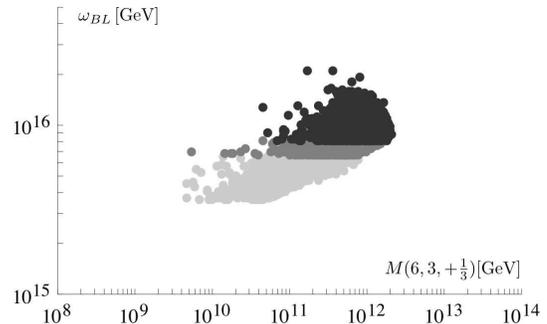}
\caption{\label{FinetunedOmegaBL63p13}$M(6,3,+\tfrac{1}{3})-\omega_{BL}$ correlation in the case of a light $(6,3,+\tfrac{1}{3})$ multiplet in the desert. Various levels of gray correspond to domains accessible for different GUT-scale limits, cf.~Section~\ref{sect:protonlimits}. $M(6,3,+\tfrac{1}{3})$ can vary only over a couple of  orders of magnitude (for the current SK limit) and the range is likely to shrink considerably in future.}
\end{figure}
\begin{figure}[ht]
\includegraphics[width=7cm]{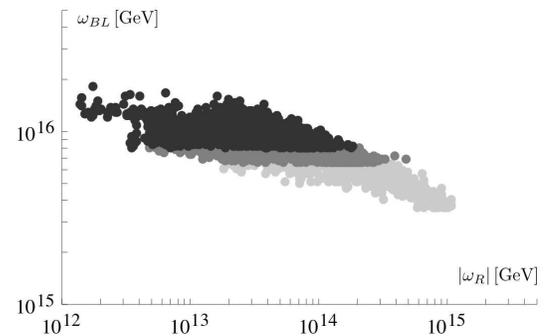}
\caption{\label{FIGGUTscale63p13}$|\omega_{R}|-\omega_{BL}$ correlation in the case of a light $(6,3,+\tfrac{1}{3})$ multiplet in the desert. The color code is the same as before, cf.~Section~\ref{sect:protonlimits}. In all of the allowed region $|\omega_{R}|\ll \omega_{BL}$ so this setting prefers an intermediate $3_c2_L2_R1_{BL}$ stage.}
\end{figure}
\begin{figure}[ht]
\includegraphics[width=7cm]{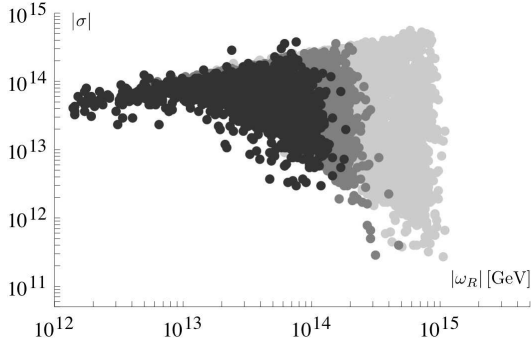}
\caption{\label{FIGOmegaRsigma63p13}$|\omega_{R}|-|\sigma|$ correlation in the case of a light $(6,3,+\tfrac{1}{3})$ multiplet in the desert. The color code is the same as before, cf.~Section~\ref{sect:protonlimits}. $B-L$ as high as almost $10^{15}$GeV can be reached for the current SK limit, with the best Hyper-K sensitivity limit the maximum lowers to few$\times10^{14}$GeV.}
\end{figure}
It is worth noticing that: 
{\it i)} In the fully consistent settings $(6,3,+\tfrac{1}{3})$ is pinned down to a relatively narrow region around $10^{11}$GeV, cf.~FIG.~\ref{FinetunedOmegaBL63p13}. 
{\it ii)} For all consistent configurations we find $|\omega_{R}|\ll \omega_{BL}$ so these settings generally prefer an intermediate $3_c2_L2_R1_{BL}$ stage, cf.~FIG.~\ref{FIGGUTscale63p13}. 
{\it iii)} There is just a little room left if the current Super-K limits get improved considerably in future, see FIG.~\ref{FinetunedOmegaBL63p13}. Moreover, since the two-loop effects tend to further lower the GUT scale with respect to the one-loop estimates (even as much as half an order of magnitude)~\cite{Bertolini:2009qj}, 
this class of scenarios may be ultimately testable at Hyper-K.
{\it iv)} Finally, the actual upper limit on the $B-L$ scale is stretched to almost $10^{15}$~GeV and it slowly decreases for stronger proton-decay limits, cf.~FIG.~\ref{FIGOmegaRsigma63p13}.  

However, one should be more careful here because these results can be  biased by the stability of the numerical approach we are using, cf. Section~\ref{sect:technicalities}. Namely, the system of equations implementing the unification constraints can be efficiently solved for the position of $(6,3,+\tfrac{1}{3})$ and for the overall shift of the spectrum if and only if $(6,3,+\tfrac{1}{3})$  is considerably lighter than the next-to-lightest threshold at play (typically a gauge boson associated to the $2_{R}1_{BL}\to 1_{Y}$ breaking); otherwise it becomes highly non-linear and, hence,  difficult to handle. However, as one can see in FIG.~\ref{Stability_63p13},  for the estimate of the upper limit on $\sigma$ this issue is less important because some of the couplings (namely, $\beta_{4}$ and $\beta_{4}'$) turn non-perturbative yet before this issue really occurs. Moreover, the  shape of the new upper limit on the $B-L$ scale is such that one is likely to miss solutions in the lower-$B-L$ regime which is not of the utmost importance here. 
\begin{figure}[ht]
\includegraphics[width=7cm]{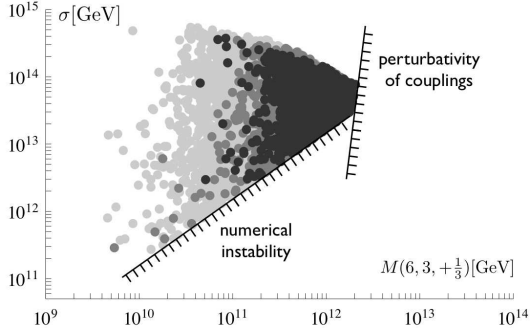}
\caption{\label{Stability_63p13}
Correlation between the mass of the $(6,3,+\tfrac{1}{3})$ threshold and the allowed $B-L$ scale $\sigma$. There are two basic stability issues that bias the estimate of the span of the allowed domains: first, there is the technical requirement we impose on the hierarchy between the lightest and next-to-lightest thresholds, i.e, $(6,3,+\tfrac{1}{3})$ and the gauge sector associated to the $3_{c}2_{L}2_{R}1_{BL}$ breaking  which cuts the parametric space from below right; for large $\sigma$'s this, however, becomes irrelevant because some of the couplings (namely, $\beta_{4}$ and $\beta_{4}'$)  become non-perturbative yet before such a numerical instability can affect the relevant upper bound.} 
\end{figure}
\paragraph{A specific example with a light $(6,3,+\tfrac{1}{3})$:}
\label{sect:specific63p13}
\begin{figure}[ht]
\includegraphics[width=7cm]{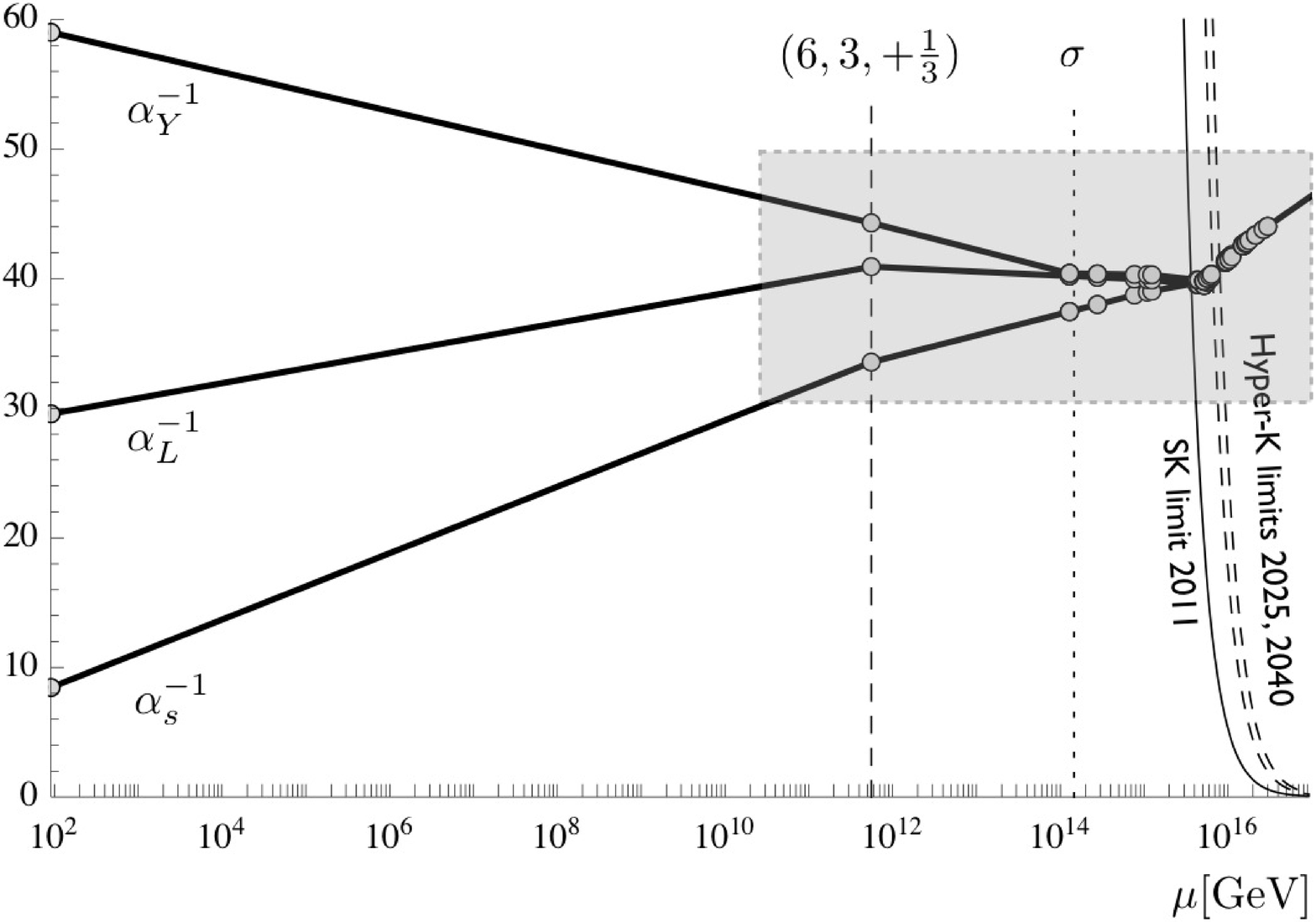}
\includegraphics[width=7cm]{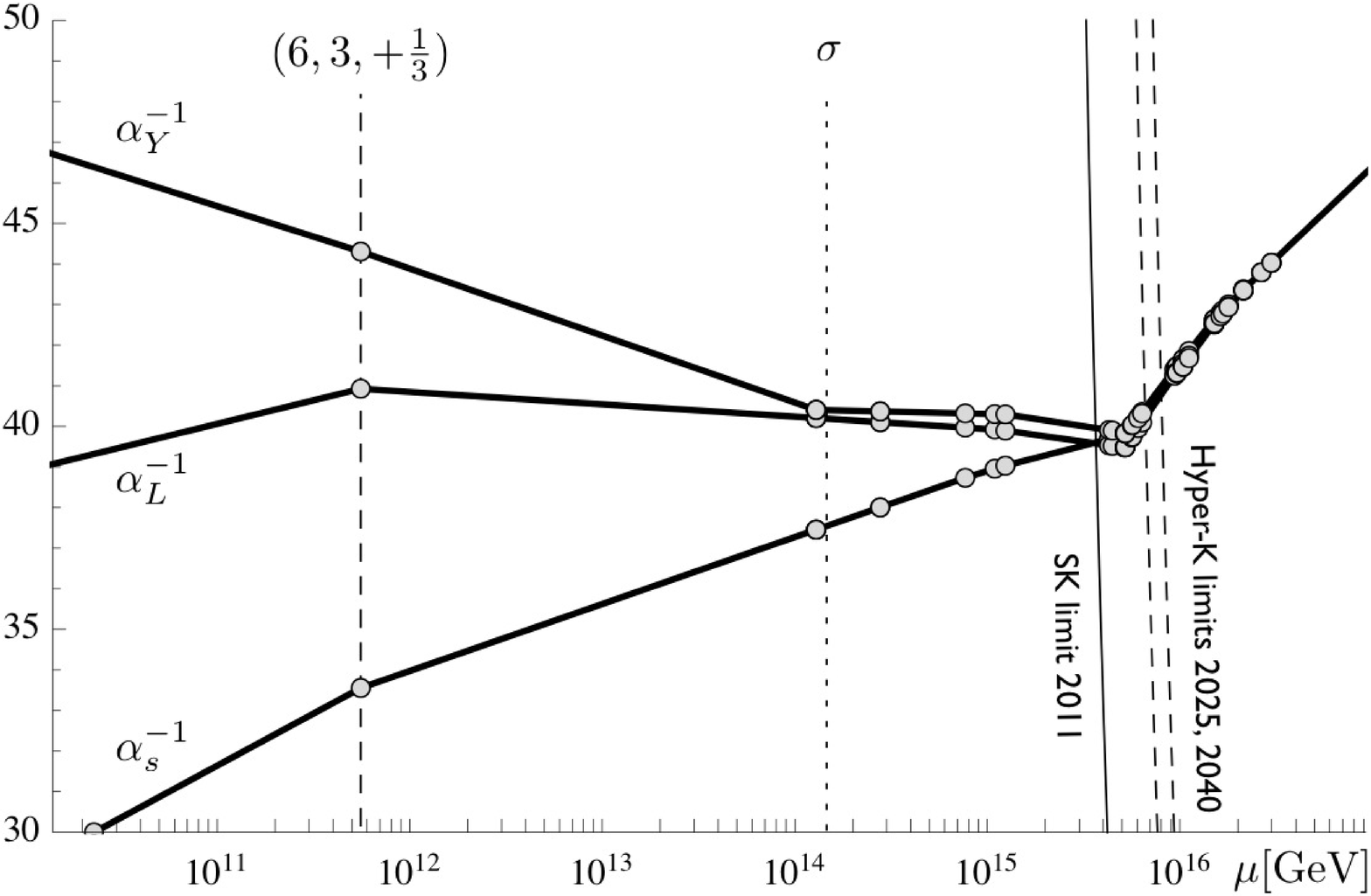}
\caption{\label{FIGrunning63p13}Unification of the effective SM gauge couplings in a sample setting with a light $(6,3,+\tfrac{1}{3})$ multiplet (here at around $5.6\times 10^{11}$GeV, cf. Section~\ref{sect:specific63p13}) with the shaded area magnified on the lower panel. The short $3_c2_L2_R1_{BL}$ stage is clearly visible here. The small circles indicate the positions of various thresholds (for details, see TABLE~\ref{sample63p13}) inflicting changes in the three curve's slopes. The almost-vertical solid and dashed lines correspond to the current and future proton-decay limits, cf.~Section \ref{sect:protonlimits}. The displayed setting is compatible (at one loop) with the current SK limit, but it can be refuted by the Hyper-Kamiokande. The dotted vertical line indicates the position of the $B-L$ scale.} 
\end{figure}
The ``effective'' SM gauge coupling evolution with a light $(6,3,+\tfrac{1}{3})$ is exemplified in FIG.~\ref{FIGrunning63p13} where the values of the input parameters as specified in the left row of TABLE~\ref{TableSampleParameters} have been used and $\tau$ is calculated so that the desired $M(6,3,+\tfrac{1}{3})=5.57\times 10^{11}$ GeV is obtained. Note that the small $|\gamma_{2}|$ region turns out to be preferred for larger values of $|\sigma|$ and that we have chosen a solution with relatively small $|\lambda_{4}|$ and $\lambda_{4}'$ just to optically improve the expected ``clustering'' of the $(3,2,+\tfrac{7}{6})$ and $(3,2,+\tfrac{1}{6})$ multiplets at around $10^{15}$~GeV (cf.~FIG.~\ref{FIGrunning63p13}) due to their common origin within $(3,2,2,+\tfrac{2}{3})$ of $3_{c}2_{L}2_{R}1_{BL}$. A more detailed information about the relevant bosonic spectrum underlying the gauge unification in this setting is given in TABLE~\ref{sample63p13} of Appendix~\ref{samplespectrum}.
\begin{table}
\begin{tabular}{c|c|c}
\hline\hline
 & light  $(6,3,+\tfrac{1}{3})$ & light $(8,2,+\tfrac{1}{2})$ \\
\hline
parameter & value  & value \\
\hline
$\omega_{R}$ [GeV] & $-2.92 \times 10^{13}$ & $-1.46 \times 10^{16}$   \\
$\omega_{BL}$ [GeV] & $8.65  \times 10^{15}$  & $-4.04 \times 10^{12}$  \\
$\sigma$ [GeV] &  $-1.46  \times 10^{14}$ & $-3.23  \times 10^{13}$ \\
$a_{0}$ &  $ 0.50 $ & $ 0.50$  \\
$\alpha$ &  $ 0.55 $ & $ 0.47 $  \\
$\beta_{4} $ &  $ 0.61 $ & $0.60 $ \\
$\beta_{4}'$ &  $ -0.41 $  & $ -0.34 $\\
$\gamma_{2}$ &  $ -0.12 $  & $  -0.01 $\\
$\lambda_{0}$ &  $ 0.95 $ & $ -0.86 $ \\
$\lambda_{2}$ &  $ 0.34  $& $ -0.14 $ \\
$\lambda_{4}$ &  $ -0.07  $& $ -0.04$\\
$\lambda_{4}' $ &  $ -0.15  $ & $ -0.07$\\
\hline
$M(\text{threshold})$ [GeV] & $5.57\times 10^{11}$ & $2.3 \times 10^{4}$ \\
\hline\hline
\end{tabular}
\caption{\label{TableSampleParameters}Parameters underpinning the two sample settings with a light $(6,3,+\tfrac{1}{3})$ (left) and a light $(8,2,+\tfrac{1}{2})$ (right), respectively. The value of the $\tau$ parameter can be obtained from the requirement that the relevant light threshold has the mass specified in the last row.}
\end{table}

\subsubsection{Consistent setting 2: light  $(8,2,+\tfrac{1}{2})$}
\paragraph{Stable vacua with a light  $(8,2,+\tfrac{1}{2})$:}
Turning our attention to the remaining option of a light $(8,2,+\tfrac{1}{2})$ in the desert it is possible to show that (for all dimensionless couplings smaller than 1 in absolute value)  there are always tachyons in the scalar spectrum outside the following domain: $|\omega_{BL}| <|\omega_R|$, $\beta_4'<0$, $a_{0}>-0.05$, $|\gamma_2|<0.6$, $|\gamma_2|<-0.8 \beta_4'$. Moreover, only one of the eigenstates  of the mass matrix~(\ref{M82p12}) can be consistently lowered. 
\paragraph{One-loop unification with a light $(8,2,+\tfrac{1}{2})$:}
Sample regions of the parametric space that support a consistent scalar spectrum and, at the same time, provide a viable gauge coupling unification, are depicted in FIGs.~\ref{FinetunedOmegaR82p12}, \ref{FIGGUTscale82p12} and \ref{FIGOmegaBLsigma82p12}.
\begin{figure}[ht]
\includegraphics[width=7cm]{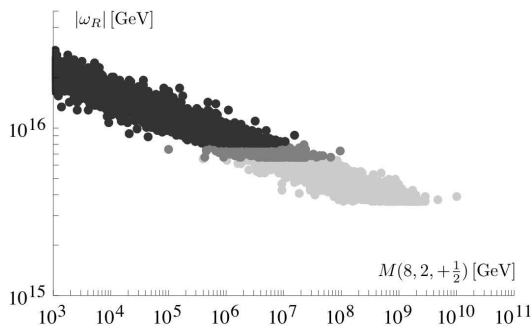}
\caption{\label{FinetunedOmegaR82p12}$M(8,2,+\tfrac{1}{2})-|\omega_{R}|$ correlation in the case of a light $(8,2,+\tfrac{1}{2})$ multiplet in the desert. The color code is the same as before, cf.~Section~\ref{sect:protonlimits}. $M(8,2,+\tfrac{1}{2})$ can vary over many orders of magnitude in the lower part of the desert, and it is  pushed down for increasing proton lifetime.}
\end{figure}
\begin{figure}[ht]
\includegraphics[width=7cm]{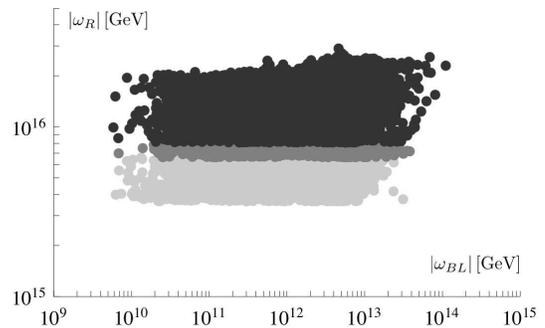}
\caption{\label{FIGGUTscale82p12}$|\omega_{BL}|-|\omega_{R}|$ correlation in the case of a light $(8,2,+\tfrac{1}{2})$ multiplet in the desert. Various levels of gray correspond to domains accessible for different GUT-scale limits, cf.~Section~\ref{sect:protonlimits}. In the whole allowed region $|\omega_{BL}|\ll \omega_{R}$ so this setting always exhibits an intermediate $4_C2_L1_R$ stage.}
\end{figure}
\begin{figure}[ht]
\includegraphics[width=7cm]{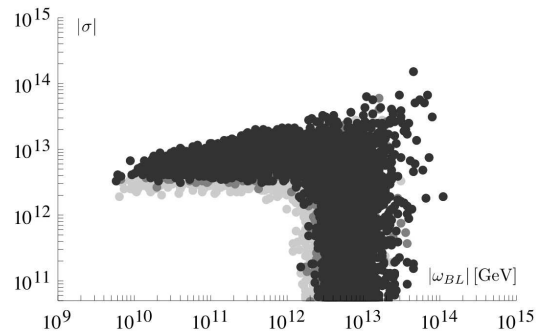}
\caption{\label{FIGOmegaBLsigma82p12}An interesting $|\omega_{BL}|-|\sigma|$ correlation in the case of a light $(8,2,+\tfrac{1}{2})$ multiplet in the desert. The color code is the same as before, cf.~Section~\ref{sect:protonlimits}. $B-L$ as high as  $10^{14}$GeV can be reached and, remarkably enough, unlike in the $(6,3,+\tfrac{1}{3})$ case the maximum reach is insensitive to the proton lifetime limit.}
\end{figure}
Note in particular that: 
{\it i)} In the fully consistent settings $(8,2,+\tfrac{1}{2})$ is narrowed down to the lower part of the desert (it is always below $10^{10}$ GeV) and even more so if proton lifetime limits get improved in the near future, cf.~FIG.~\ref{FinetunedOmegaR82p12}. Nevertheless, the allowed domain for the light threshold is much wider than in the previous case, see FIG.~\ref{FinetunedOmegaBL63p13}; hence, this scenario is likely to be more robust to the changes inflicted by two-loop effects.
In the extreme case this class of models requires $(8,2,+\tfrac{1}{2})$ close to the EW scale with possibly interesting collider effects.  
{\it ii)} For all consistent configurations we find $|\omega_{BL}|\ll |\omega_{R}|$ so these settings generally prefer an intermediate $4_C2_L1_R$ stage, cf.~FIG.~\ref{FIGGUTscale82p12}. 
{\it iii)}
The upper limit on the $B-L$ scale is pushed to about $10^{14}$~GeV and, unlike in the previous case, it is rather insensitive to possible future improvements of the proton lifetime limits, cf.~FIG.~\ref{FIGOmegaBLsigma82p12}.

Note also that there is no problem with the numerical stability here, cf. Section~\ref{sect:technicalities}, because the gap between the mass of $(8,2,+\tfrac{1}{2})$ and the next-to-lightest threshold preferred by the unification constraints is huge.
\paragraph{A specific example with a light $(8,2,+\tfrac{1}{2})$:
\label{sect:specific82p12}}
\begin{figure}[ht]
\includegraphics[width=7cm]{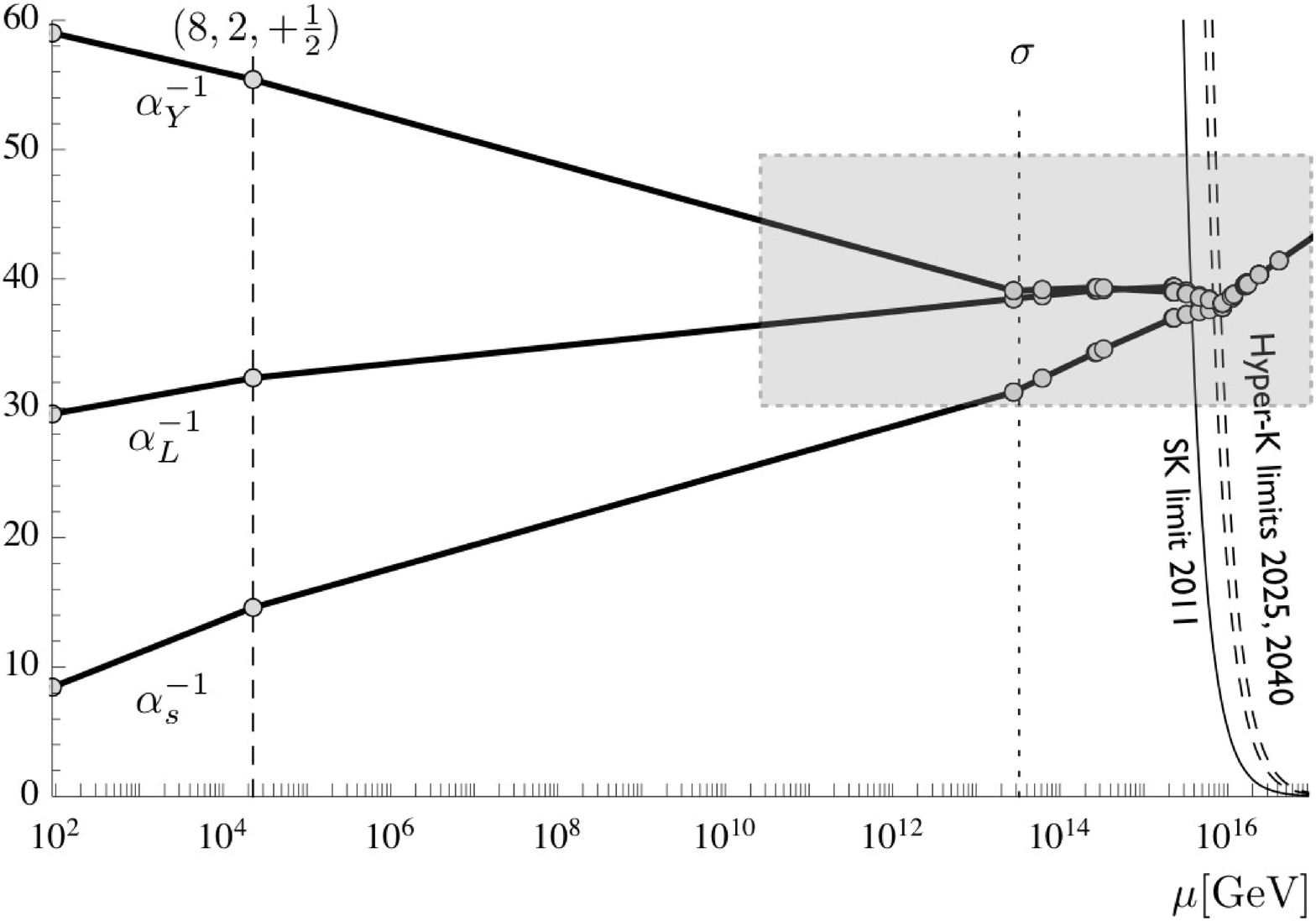}
\includegraphics[width=7cm]{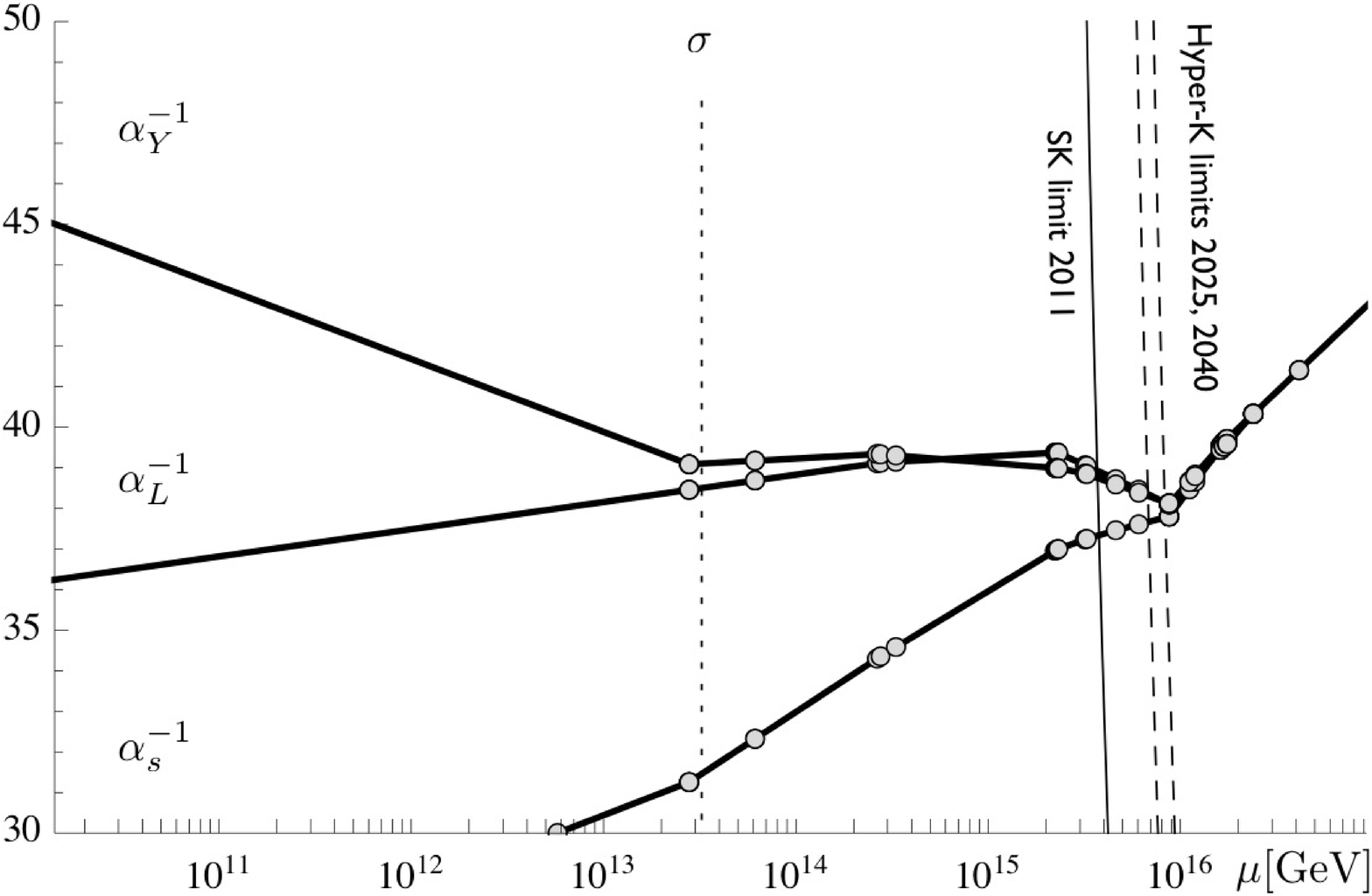}
\caption{\label{FIGrunning82p12}The same as in FIG.~\ref{FIGrunning63p13}, here with a light $(8,2,+\tfrac{1}{2})$ multiplet (at around $2.3\times 10^{4}$ GeV, 
cf.~Section~\ref{sect:specific82p12}) instead of $(6,3,+\tfrac{1}{3})$;  for details of the spectrum see TABLE~\ref{sample82p12}.  The short $4_C2_L1_R$ stage as well as the ``fake unification'' feature is clearly visible here.   The displayed setting is compatible (at one-loop) with the current SK as well as possible future Hyper-Kamiokande limits.}
\end{figure}
The ``effective'' SM gauge coupling evolution with a light $(8,2,+\tfrac{1}{2})$ is exemplified in FIG.~\ref{FIGrunning82p12} where the values of the input parameters as specified in the right column of TABLE~\ref{TableSampleParameters} have been used
and $\tau$ is fixed so that $M(8,2,+\tfrac{1}{2})=2.3\times 10^{4}$~GeV. Note that the small $|\gamma_{2}|$ region turns out to be preferred for larger values of $|\sigma|$ and we have chosen smaller $|\lambda_{4}|$ and $\lambda_{4}'$ just to optically improve the expected ``clustering'' of the remnant of the $(15,2,+\tfrac{1}{2})$ multiplet of $4_{C}2_{L}1_{R}$ (where the light $(8,2,+\tfrac{1}{2})$ comes from)  at around $3\times 10^{14}$ GeV, cf.~FIG.~\ref{FIGrunning82p12}. A more detailed information about the relevant bosonic spectrum underlying the gauge unification in this setting is given in TABLE~\ref{sample82p12} of Appendix~\ref{samplespectrum}.

\subsubsection{Further remarks}
It is perhaps interesting to note that neither $(6,3,+\tfrac{1}{3})$ nor $(8,2,+\tfrac{1}{2})$, even if pushed well below the GUT scale, are any way problematic for either BBN or $d>6$ proton-decay, cf.~Sects.~\ref{d7protondecay} and~\ref{sect:BBNconstraints}. 
As for the former, their direct coupling to quarks and leptons (by means of the $16_{F}16_{F}126_{H}^{*}$ $SO(10)$ Yukawa interaction) makes both of them decay fast enough via a simple tree-level diagram.
Similarly, neither $(6,3,+\tfrac{1}{3})$ nor $(8,2,+\tfrac{1}{2})$ can 
mediate the effective $d=7$ ($B+L$  conserving) proton decay. 

\section{The minimal  $SO(10)$ GUT revived}
\label{sect:newminimalSO10}
The previous analysis reveals several regions of the parametric space of the non-SUSY $SO(10)$ Higgs model based on the reducible representation $45_H \oplus 126_H$ that can consistently support $SO(10)\to$~SM symmetry breaking chains compatible with the electroweak data and the current proton decay limits and, simultaneously, admit for a large-enough $B-L$ breaking scale for a natural implementation of a renormalizable seesaw. 
Hence, this simple Higgs model is ready to be upgraded to a full-featured, potentially realistic and predictive $SO(10)$~GUT.   

In doing so, the central question to be addressed before approaching any of the ultimate goals of such a programme (e.g., a detailed prediction of the proton lifetime and the relevant branching ratios) is the structure of the Yukawa sector. 
\subsection{Yukawa sector of the minimal $SO(10)$ GUTs}
It is easy to see that the Higgs model containing just $45_H$ and $126_H$ can not, at renormalizable level, support a viable Yukawa sector as there is only one contraction available in such a case, namely, $16_{F}f^{126}16_{F}126_{H}^{*}$. Hence, the flavour structure is entirely governed by a single (symmetric) matrix of Yukawa couplings $f^{126}$ and no mixing nor featured fermionic spectra can be generated.

The minimal potentially realistic extension of the $45_{H}\oplus 126_{H}$ setting amounts to adding an extra $10$- or $120$-dimensional representation which can smear the degeneracy of the effective Yukawa matrices across different fermionic species; for a more detailed discussion see, e.g.,~\cite{Babu:1992ia} or, more recently,~\cite{Bajc:2005zf}. In this respect, it is interesting to quote namely the results of the new numerical analysis~\cite{Joshipura:2011nn} attempting to fit the  SM flavour structure onto the effective mass matrices emerging in both the $126_{H}\oplus 10_{H}$ as well as the $126_{H}\oplus 120_{H}$ cases: Interestingly, the former option is strongly preferred and, moreover, successful fits require a dominance of the type-I seesaw contribution\footnote{This feature is closely related to the need to avoid the $b$-$\tau$ Yukawa unification in the non-SUSY settings which, however, is generically favoured by type-II seesaw.}. However, as interesting as these results are, they are still not entirely decisive as there are various sources of uncertainties\footnote{In particular: {\it i)} the weights of the SM-doublet VEVs entering the relevant sum-rules, cf. Eqs.~(\ref{sum-rules}), were taken  uncorrelated, {\it ii)} the running fermionic masses were extrapolated to the GUT-scale vicinity under the bold assumption of no thresholds in the desert and {\it iii)} higher order corrections to the relevant sum-rules inherent to non-SUSY settings were not taken into account.} that have not been taken into account in~\cite{Joshipura:2011nn}.  

Nevertheless, the Higgs sector based on $45_{H} \oplus 10_{H} \oplus 126_{H}$ is clearly the first choice; not only it has a better chance to be compatible with the fermionic data, but the addition of an extra $10_{H}$ rather than a larger multiplet like $120_{H}$ only minimally disturbs the results obtained in the previous parts, see also the comments in Sect.~\ref{sect:technicalities}.

For the sake of completeness, let us reiterate the Yukawa-sector sum-rules relevant to this setting. In full generality, one can write a renormalizable Lagrangian density\footnote{Note that $10^{*}$ of $SO(10)$ is equivalent (in the representation sense) to $10$, so both $16^{2}10$ and $16^{2}10^{*}$ are allowed in non-SUSY scenarios.}
\be\label{full}
{\cal L}  \ni   {16_{F}}(f_{1}^{10}{10_{H}}+f_{2}^{10}10_{H}^{*}+  f^{126}{{126}_{H}^{*}}){16_{F}} +{\rm h.c.}\,,
\ee
which is parametrized by three complex symmetric matrices $f_{1,2}^{10}$ and $f^{126}$. It leads to the following general tree-level formulae for the effective SM quark and lepton  mass matrices\footnote{Let us remind the reader that a good grip on the Yukawa couplings is also necessary for a reliable account of the $d=6$ proton-decay amplitudes because they depend on the matrix elements of the unitary transformation bringing the quarks and leptons from the  current to the mass basis.}
\begin{align}
M_{u} & =  Y_{1}^{10}v_{u}^{10}+Y_{2}^{10}v_{d}^{10 *}+Y^{126}v_{u}^{126}\,, \nn \\
M_{d} & =  Y_{1}^{10}v_{d}^{10}+Y_{2}^{10}v_{u}^{10 *}+Y^{126}v_{d}^{126}\,, \label{sum-rules} \\
M_{\ell} & =  Y_{1}^{10}v_{d}^{10}+Y_{2}^{10}v_{u}^{10 *} -3 Y^{126}v_{d}^{126}\,, \nn \\
M^{\rm D}_{\nu} & =  Y_{1}^{10}v_{u}^{10}+Y_{2}^{10}v_{d}^{10 *}-3 Y^{126}v_{u}^{126}\,, \nn\\
M^{\rm M,I}_{\nu}& =c^{\rm I}Y^{126}\sigma, \quad M^{\rm M,II}_{\nu} =c^{\rm II}Y^{126}w^{126},\nn
\end{align}
where $Y_{1,2}^{10}$ and $Y^{126}$ are proportional to the corresponding $f$-matrices in (\ref{full}), the subscripts D and M denote the Dirac and Majorana segments of the  neutrino mass matrix and $c^{\rm I,II}$ (for type-I and type-II seesaw, respectively) combine various extra numerical factors such as the relevant Clebsch-Gordan coefficients. Let us note that, unlike in SUSY, here the $10_H$ of $SO(10)$ can be populated by real components~\cite{Babu:1984mz}, which would further reduce the number of independent couplings -- indeed, in such a case, the second term in (\ref{full}) would be just a repetition of the first one. However, this setting is pathological as it leads to a GUT-scale near-equality of the $b$- and $t$-quark masses, cf.~\cite{Bajc:2005zf}. 

It should be stressed that in the full model, the projections of the SM Higgs VEVs onto the indicated components of the relevant $SO(10)$ multiplets 
$v_{u}^{10} = \langle(1,2,+\tfrac{1}{2})_{10}\rangle$, $v_{d}^{10} = \langle(1,2,-\tfrac{1}{2})_{10}\rangle$, 
$v_{u}^{126} = \langle(1,2,+\tfrac{1}{2})_{126^{*}}\rangle$ and 
$v_{d}^{126} = \langle(1,2,-\tfrac{1}{2})_{126^{*}}\rangle$ as well as the VEV of the type-II $SU(2)_{L}$ triplet $w^{126}=\langle(1,3,+1)_{126^{*}}\rangle$ are calculable functions of the parameters entering the scalar potential. 
Note also that, in full generality, the formulae (\ref{sum-rules}) correspond to the two-Higgs-doublet realization of the SM Higgs sector; assuming that only one of the doublets survives down to the electroweak scale (i.e., implementing a single fine-tuning in the relevant generalization of the mass matrix~(\ref{matrix:12p12})) one should further assume either\footnote{Actually, only the former option has a chance to work in practice because the latter immediately implies an apparently wrong relation $M_{d}=M_{\ell}$ so we discard it.} $v_{u}^{10}=v_{u}^{126}=0$ or $v_{d}^{10}=v_{d}^{126}=0$.

In connection to this, one should mention a couple of other interesting features inherent to the models with $45_{H}$ that have no counterpart in many other settings including the popular variant with the GUT symmetry broken by $54_{H}$~\cite{Lazarides:1980nt,Buccella:1984ft,Buccella:1986hn}: {\it i)} First, in the former case, a significant admixture of both the $10_{H}$ and $126_{H}$ components within the SM Higgs doublet, which turns out to be essential for realistic fits of the Yukawa system (\ref{sum-rules}) to the quark and lepton data, is naturally obtained via the direct mixing term $10_H 126^*_H 45_H 45_H$. In the latter case, however, there is no  such a term and the mixing is governed namely by the $10_H 126^*_H 126_H 126_H$ contraction which, however, yields and extra suppression of the order of $M_{B-L}^{2}/M_{G}^{2}$.
{\it ii)} Second, the settings in which the GUT symmetry is broken by the VEVs of $45_{H}$ generally feature an almost-automatic suppression of the type-II seesaw which, as mentioned previously, is not only welcome due to the generic GUT-scale non-equality of the $b$ and $\tau$ Yukawas in non-SUSY settings~\cite{Xing:2007fb}, but it seems to be even crucial for successful Yukawa fits, cf.~\cite{Joshipura:2011nn}.
Indeed, on general grounds, one expects that 
in theories in which the $D$-parity\footnote{$D$-parity is a discrete symmetry acting as a charge conjugation in the left-right
symmetric context~\cite{Kuzmin:1980yp,Kibble:1982dd}, 
and, as such, it plays the role of a left-right symmetry (enforcing, for instance, equal $SU(2)_{L}$ and $SU(2)_{R}$ gauge
couplings). As a part of the $SO(10)$ algebra, it is a good symmetry until it gets broken by either a $D$-odd singlet in $45_{H}$ (or $210_{H}$) or by any $SU(2)_{R}$-breaking VEV, thus allowing for a left-right asymmetric scalar 
spectrum~\cite{Chang:1983fu,Chang:1984uy}.} is broken before the $SU(2)_{R}$ symmetry, the type-II contribution to the light neutrino masses is naturally suppressed by a factor of $M_{B-L}^{2}/M_G^2$ with respect to the 
type-I term~\cite{Chang:1985en}.
Again, this is not the case in models based on the $54_{H}$ where the D-parity is preserved down to the $SU(2)_{R}$-breaking scale and, thus, no extra suppression of type-II seesaw occurs. 

\subsection{Predictivity and testability}
Concerning the predictivity of the renormalizable model based on $45_{H} \oplus 10_{H} \oplus 126_{H}$ in the Higgs sector, there are several aspects worth a comment here.

{\it i)} Yukawa sector complexity: There are in general three independent complex symmetric matrices entering the effective sum-rules~(\ref{sum-rules}), to be compared to just two such structures encountered in, e.g., the minimal potentially realistic Yukawa sector in supersymmetry~\cite{Clark:1982ai}. This, however, {\it does not} necessarily imply a loss of predictivity in the Yukawa sector: First, the weights of the SM VEVs entering Eqs.~(\ref{sum-rules}) are in general stronger correlated here than in the minimal SUSY case (cf.~\cite{Bajc:2004xe}) because here the doublet mass matrix~(\ref{matrix:12p12}) is lower-dimensional. Second, with only one doublet pushed down to the electroweak scale, the system~(\ref{sum-rules}) is simplified and the correlations among different species become much tighter. This is also well reflected by the preliminary results of a dedicated numerical analysis~\cite{Bertolini:2012ip}. 

{\it ii)} Vacuum stability: Unlike in the (global) SUSY case where the positivity of the scalar mass-squares is automatic in any SUSY-preserving vacuum, the consistency requirements here narrow the potentially interesting domains of the parametric space down to just few small patches (for instance those identified in Sect.~\ref{sect:unification}). On the other hand, given the higher number of contractions available in the non-SUSY case even at the renormalizable level (to be compared to just several such terms entering the Higgs superpotential in SUSY) the set of SM-like vacua is clearly higher-dimensional (see the number of parameters in TABLE~\ref{TableSampleParameters} versus a single complex parameter in SUSY, cf.~\cite{Bajc:2004xe}).

{\it iii)} Radiative corrections: Unlike many popular SUSY $SO(10)$ variants, cf. \cite{Aulakh:2003kg,Aulakh:2006hs} the model under consideration is asymptotically free (with $b= -12$) so it remains weakly coupled up to the Planck scale. On the other hand, its radiative structure is much more involved than that of the simplest SUSY scenarios and consistent calculations are technically much more demanding. In this respect let us reiterate that the nature of the problem calls for a two-loop RGE analysis based on a detailed knowledge of the one-loop spectrum and in this work we have just performed the first steps in that direction.
\vskip 1mm
Hence, without a detailed analysis it is rather difficult to assess the predictive power of the model under consideration. Nevertheless, even the first results obtained in Sect.~\ref{sect:unification} indicate that the up-coming large volume experiments such as Hyper-K can impose very strong cuts to its (already rather constrained) parametric space, possibly covering the entire remaining volume. 

\vskip 2mm
Sometimes, it is suggested to further enhance the Yukawa-sector predictivity of the non-SUSY models by imposing an extra global $U(1)$ symmetry of the Peccei-Quinn (PQ) type~\cite{Peccei:1977hh,Peccei:1977ur} which, if it transforms $10_{H}$ non-trivially, forbids one of the $f^{10}_{1,2}$ couplings in the Lagrangian~(\ref{full}). Since, in that case, also $126_{H}$ would have to be PQ-charged, such a symmetry would be broken at the same scale as $U(1)_{B-L}$, thus linking the PQ symmetry-breaking scale to the neutrino masses. In this respect, it is very interesting that a seesaw scale in the preferred $10^{13\div 14}$ GeV ballpark is indeed very close to the $10^{9 \div 12} \ \text{GeV}$ PQ-symmetry-breaking window favoured by astrophysics and cosmology (see e.g.~\cite{Raffelt:2006cw}) and there are several attempts in the literature to construct a viable unified model along these lines (see, e.g,~\cite{Mohapatra:1982tc}, or more recently~\cite{Bajc:2005zf}). On the other hand, since $
\vev{126_{H}}$ can not break the rank of $SO(10) \otimes U(1)_{PQ}$ by more than a single unit, a global linear combination of $U(1)_{PQ}$, $U(1)_{R}$ and $U(1)_{B-L}$ survives down to the electroweak scale and only there it gets finally broken by the electroweak doublet(s); this, however, is unacceptable as the EW-scale PQ-symmetry breaking gives rise to an easily visible axion~\cite{Weinberg:1977ma,Wilczek:1977pj}. 
Thus, a consistent implementation of this interesting scheme calls for a further complication of the Higgs sector, which we shall not entertain here. 

\section{Conclusions and outlook}

In this work we have been concerned with a class of simple renormalizable $SO(10)$ Higgs models in which the first stage of symmetry breaking is triggered by the 45-dimensional adjoint Higgs representation. These settings, discarded a long ago due to inherent tree-level tachyonic instabilities developing in most of the physically interesting scenarios, have been recently shown to be revived by quantum effects. However, many important aspects of these scenarios, such as, for instance, the allowed ranges for the unification as well as various intermediate scales, as important as these are for any realistic model building, were never studied in sufficient detail to allow for a qualified assessment of their physical relevance. 

Focusing on the variant with $45_{H}\oplus 126_{H}$ in the Higgs sector, we worked out the complete tree-level spectrum and, with such an extra information at hand, performed a simple analysis of the gauge-unification patterns. Unlike in the previous studies based on the minimal-survival hypothesis, that show a no-go for scenarios with the $B-L$ scale above $10^{10}$~GeV~\cite{Bertolini:2009qj}, we found several domains in the parametric space of these models that can support a consistent gauge-unification with $B-L$ as high as $10^{14}$~GeV without encountering any tachyonic instabilities or proton lifetime issues. The key to this unexpected behaviour is an accidentally light threshold in the desert which affects the gauge-unification picture in a suitable way. 
We identified two distinct classes of such viable solutions: in the first case, an intermediate-scale multiplet transforming as $(6,3,+\tfrac{1}{3})$ of the SM supports $SO(10)\to$~SM descents featuring a short $SU(3)_{c}\otimes SU(2)_{L}\otimes SU(2)_{R}\otimes U(1)_{B-L}$ intermediate-symmetry stage, while the second option including a relatively light $(8,2,+\tfrac{1}{2})$ supports $SO(10)$ breaking chains passing through the $SU(4)_{C}\otimes SU(2)_{L}\otimes U(1)_{R}$ intermediate symmetry.
Remarkably enough, in all the cases of interest, the unification scale turns out to be rather close to the current Super-Kamiokande proton-lifetime lower bound.

This, however, opens up an intriguing possibility to construct a simple, renormalizable and testable $SO(10)$ GUT with $45_{H}\oplus 126_{H}\oplus 10_{H}$ in the Higgs sector which, in view of the recent failure of the simplest supersymmetric $SO(10)$ model~\cite{Aulakh:2005mw,Bertolini:2006pe}, can even be viewed as the new minimal potentially realistic $SO(10)$ GUT.

\vskip 2mm
Nevertheless, this study provides only the first glimpse on the ultimate viability of such a framework and there is much more still to be done. Let us reiterate that simple non-SUSY models suffering from significant tree-level vacuum instabilities generically call for a refined two-loop RGE approach (assuming one-loop scalar spectrum) because only in such a case the tachyons are really under control. In this respect, the results of the current analysis, taking into account only the minimal set of radiative corrections necessary for the scalar spectrum regularization, can quantitatively (though not qualitatively) differ from those to be obtained in a future full one-loop effective potential analysis.

Remarkably enough, extrapolating the relative size and direction of the two-loop effects observed in~\cite{Bertolini:2009qj} to the current scheme, the chances for its ultimate testability at future experiments look rather promising. Indeed, any further significant decrease of the maximum allowed unification scale due to two-loop effects would allow the up-coming large-volume facilities such as Hyper-Kamiokande to scan over the full physically interesting domain in the parametric space of this class of models.

\subsection*{Acknowledgments}

S.B. is partially supported by MIUR and the EU UNILHC-grant agreement PITN-GA-2009-237920. The work of L.DL. was supported by the DFG through the SFB/TR 
9 ``Computational Particle Physics''. The work of M.M. is supported by the Marie Curie Intra European Fellowship
within the 7th European Community Framework Programme
FP7-PEOPLE-2009-IEF, contract number PIEF-GA-2009-253119, by the EU
Network grant UNILHC PITN-GA-2009-237920, by the Spanish MICINN
grants FPA2008-00319/FPA and MULTIDARK CAD2009-00064
(Consolider-Ingenio 2010 Programme) and by the Generalitat
Valenciana grant Prometeo/2009/091.

\appendix
\section{Details of the scalar potential\label{app:details}}

The 2-index and 5-index completely antisymmetric tensors of $SO(10)$ are labelled respectively as $\phi_{ij}$ and $\phi_{ijklm}$. 
Given the dual map 
\be
\tilde{\phi}_{ijklm} = - \frac{i}{5!} \epsilon_{ijklmnopqr} \phi_{nopqr} \, ,
\ee
we can define the self-dual and the anti-self-dual irreducible components of $\phi_{ijklm}$ as
\bea
\label{defsigma}
\Sigma_{ijklm} &=& \frac{1}{\sqrt{2}} \left( \phi_{ijklm} + \tilde{\phi}_{ijklm}  \right) \, , \\ 
\label{defsigmabar}
\Sigma^*_{ijklm} &=& \frac{1}{\sqrt{2}} \left( \phi_{ijklm} - \tilde{\phi}_{ijklm}  \right) \, .  
\eea
Then the relevant contractions in the scalar potential of~\eq{scalpotgen} are given by  
\bea
&& (\phi \phi)_0  \equiv \phi_{ij} \phi_{ij}, \;\; (\Sigma \Sigma^*)_0 \equiv \Sigma_{ijklm} \Sigma^*_{ijklm}  \\ \nn \\
&& (\phi \phi)_0 (\phi \phi)_0  \equiv \phi_{ij} \phi_{ij} \phi_{kl} \phi_{kl} \\ \nn \\
&& (\phi \phi)_2 (\phi \phi)_2  \equiv \phi_{ij} \phi_{ik} \phi_{lj} \phi_{lk} \nn
\eea
\bea
&& (\Sigma \Sigma^*)_0 (\Sigma \Sigma^*)_0  \equiv \Sigma_{ijklm} \Sigma^*_{ijklm} \Sigma_{nopqr} \Sigma^*_{nopqr} \nn \\ \nn \\
&& (\Sigma \Sigma^*)_2 (\Sigma \Sigma^*)_2  \equiv \Sigma_{ijklm} \Sigma^*_{ijkln} \Sigma_{opqrm} \Sigma^*_{opqrn} \nn \\ \nn \\
&& (\Sigma \Sigma^*)_4 (\Sigma \Sigma^*)_4  \equiv \Sigma_{ijklm} \Sigma^*_{ijkno} \Sigma_{pqrlm} \Sigma^*_{pqrno} \\ \nn \\
&& (\Sigma \Sigma^*)_{4'} (\Sigma \Sigma^*)_{4'}  \equiv \Sigma_{ijklm} \Sigma^*_{ijkno} \Sigma_{pqrln} \Sigma^*_{pqrmo} \nn \\ \nn \\
&& (\Sigma \Sigma)_2 (\Sigma \Sigma)_2  \equiv \Sigma_{ijklm} \Sigma_{ijkln} \Sigma_{opqrm} \Sigma_{opqrn} \nn \\ \nn \\
&& (\Sigma^* \Sigma^*)_2 (\Sigma^* \Sigma^*)_2  \equiv \Sigma^*_{ijklm} \Sigma^*_{ijkln} \Sigma^*_{opqrm} \Sigma^*_{opqrn} \nn
\eea
\bea
&& (\phi)_2 (\Sigma \Sigma^*)_2 \equiv \phi_{ij} \Sigma_{klmni} \Sigma^*_{klmnj} \nn \\ \nn \\
&& (\phi \phi)_0 (\Sigma \Sigma^*)_0 \equiv \phi_{ij} \phi_{ij} \Sigma_{klmno} \Sigma^*_{klmno} \nn \\ \nn \\
&& (\phi \phi)_4 (\Sigma \Sigma^*)_4 \equiv \phi_{ij} \phi_{kl} \Sigma_{mnoij} \Sigma^*_{mnokl}  \\ \nn \\
&& (\phi \phi)_{4'} (\Sigma \Sigma^*)_{4'} \equiv \phi_{ij} \phi_{kl} \Sigma_{mnoik} \Sigma^*_{mnojl} \nn \\ \nn \\
&& (\phi \phi)_2 (\Sigma \Sigma)_2 \equiv \phi_{ij} \phi_{ik} \Sigma_{lmnoj} \Sigma_{lmnok} \nn \\ \nn \\
&& (\phi \phi)_2 (\Sigma^* \Sigma^*)_2 \equiv \phi_{ij} \phi_{ik} \Sigma^*_{lmnoj} \Sigma^*_{lmnok} \, . \nn
\eea
We have checked that this constitutes a complete set of $SO(10)$ invariants for the $45$-$126$ system at the renormalizable level.

\section{The scalar spectrum ($45_H \oplus 126_H$)\label{app:tree_spectrum}}
\subsection{Vacuum manifold and stationarity conditions}
\label{statconds}

The scalar potential \eq{scalpotgen} evaluated on the SM vacuum parametrized by $\omega_{BL}$, 
$\omega_{R}$ and $\sigma$, cf. \eq{vevs} reads
\bea
\vev{V} && =
-\mu ^2 \left(3 \omega_{BL}^2+2 \omega_R^2\right) 
+a_0 \left(12 \omega _R^2 \omega_{BL}^2+9 \omega_{BL}^4+4 \omega _R^4\right) \nn \\ \nn \\
&& + \frac{a_2}{2} \left(3 \omega_{BL}^4+2 \omega _R^4\right)-2 \nu ^2 |\sigma|^2+4 \lambda _0 |\sigma|^4 \nn \\ \nn \\
&& +2 \tau  \left(3 \omega_{BL}+2 \omega _R\right) |\sigma|^2
+2\alpha  \left(3 \omega_{BL}^2+2 \omega _R^2\right) |\sigma|^2 \nn \\ \nn \\
&& -4\beta'_{4} \left(6 \omega _R\omega_{BL}+3 \omega_{BL}^2+ \omega _R^2\right) |\sigma|^2 \, .
\eea
It is perhaps worth noting that not all couplings in expressions (\ref{V45})--(\ref{V45126}) are present here; the reason is the absence of suitable terms quartic in the available VEVs in some of the contractions. As an example consider $\eta_2$, the coefficient of the $126^4_H$ contraction in Eq.~(\ref{V126}), which enters neither the 
vacuum manifold nor the stationary conditions or the tree-level spectrum. 
This can be understood by looking at the decomposition of the  
the relevant invariant under $SU(5) \otimes U(1)_Z$ which never contains more than a single submultiplet $(1,+10)$ that is the only component of 
$126_H$ that can receive a SM-preserving VEV (recall that $\vev{126_H}$ preserves $SU(5)$).  Indeed, one has at best $126_H^4 \supset (1,+10) (\overline{15},-6) (\overline{15},-6) (\overline{50},+2)$, i.e., three derivatives are needed in order for $\eta_{2}$ to enter anywhere. A similar reasoning can be applied to the other couplings, hence fully justifying their presence/absence within all the relevant structures. However, most of  such couplings (e.g.,  $\lambda_{2}$, $\lambda_{4}$ etc.) reappear in the tree-level broken-phase mass matrices and, ultimately, all of them appear at the full one-loop effective potential level.

The corresponding stationary equations can be conveniently rewritten as 
\begin{widetext}
\bea
\label{stateq45_1}
&& \frac{1}{6 (\omega_{BL} - \omega_R)} \left( \frac{\partial \vev{V}}{\partial \omega_{BL}} - \frac{3}{2} \frac{\partial \vev{V}}{\partial \omega_{R}} \right) = -\mu ^2+ a_0 \left(6 \omega_{BL}^2+4 \omega _R^2\right)+a_2 \left(\omega _R \omega_{BL}+ \omega_{BL}^2+ \omega _R^2\right) 
+2 \alpha |\sigma|^2 +2 \beta'_{4} |\sigma|^2,  \\ \nn\\
&&
\label{stateq45_2}\frac{1}{4 (\omega_{BL} - \omega_R)} \left( \frac{2}{3} \omega_R \frac{\partial \vev{V}}{\partial \omega_{BL}}  - \omega_{BL} \frac{\partial \vev{V}}{\partial \omega_{R}} \right) = 
a_2 \omega _R \omega_{BL} \left(\omega_{BL}+\omega _R\right)
 - \tau |\sigma|^2 + 2 \beta'_{4} \left(3  \omega_{BL} + 2 \omega _R  \right) |\sigma|^2 , \\ \nn \\
\label{stateqsigma}
&& \frac{\partial \vev{V}}{\partial \sigma} = 2 \sigma^* \Bigl[-\nu ^2 +4 \lambda _0 |\sigma|^2 +\tau \left( 3 \omega_{BL} +2 \omega _R \right)
+\alpha  \left( 3  \omega_{BL}^2+2 \omega _R^2 \right)
-2 \beta'_{4} \left( 6  \omega _R \omega_{BL} +3 \omega_{BL}^2 + \omega _R^2 \right) \Bigr] \, ,
\eea
\end{widetext}
which hold away from the standard $SU(5) \otimes U(1)_Z$ vacuum ($\omega_{BL} = \omega_R$).


\subsection{Tree-level scalar spectrum in the SM limit}
\label{scalspectSM}

Let us label the scalar states with respect to the $SU(3)_c \otimes SU(2)_L \otimes U(1)_Y$ algebra. 
Applying first the stationary conditions in~\eq{stateq45_1} and~\eq{stateqsigma} one finds:
\subsubsection{States with components in $45_H$ only}
Besides the classical pair of the would-be tachyons with mass formulae
\bea
\label{M2130}
M^2 (1,3,0) &=& - 2 a_2 (\omega_{BL} - \omega_R) (\omega_{BL} + 2 \omega_R) \, , \\
\label{M2810}
M^2 (8,1,0) &=& - 2 a_2 (\omega_R - \omega_{BL}) (\omega_R + 2 \omega_{BL}) \, ,
\eea
in this sector one can identify 12 Goldstone boson modes with
\be\label{Goldstones1}
M^2 (3,2,-\tfrac{5}{6})  =  0 \, . \\
\ee
The remaining components of $45_H$ mix with those in $126_H$ and will be discussed below. 

\subsubsection{States with components in $126_H$ only}
Starting with pure states with components in $126_H$ one has:
\begin{widetext}
\begin{align}
& M^2 (1,1,+2) = 2 \left( 4 \lambda_2 + 3 \lambda_4 + 16 \lambda'_{4} \right) |\sigma|^2 
- 4 \omega _R \left(\tau -6 \beta'_4 \omega_{BL}\right) \, , \nn \\ 
& M^2 (1,3,-1) = 
8 \left( 2 \lambda_2 + 3 \lambda_4 + 2 \lambda'_{4} \right) |\sigma|^2
- 2 \left(3 \omega_{BL}+\omega _R\right) \left(\tau -2 \beta'_4 \omega_R\right) \, , \nn \\ 
& M^2 (\overline{3},1,+\tfrac{4}{3}) = 
4 \left( 3 \lambda_2  + 3 \lambda_4 + 4 \lambda'_{4} \right) |\sigma|^2
+ 2 \left(
\omega_{BL} \left(4 \beta'_4 \left(\omega_{BL}+2 \omega _R\right)+\beta _4 \omega_{BL}\right)
- \tau  \left(\omega_{BL}+2 \omega_R\right) \right) \, , \nn \\
& M^2 (3,3,-\tfrac{1}{3}) =
4 \left( 3 \lambda_2  + 3 \lambda_4 + 4 \lambda'_{4} \right) |\sigma|^2 
- 2 \tau  \left(2 \omega_{BL}+\omega _R\right) + 2 \beta _4 \omega_{BL}^2
+ 4 \beta'_4 \left(3 \omega _R \omega_{BL}+2 \omega_{BL}^2+\omega _R^2\right)
 \, , \\ 
& M^2 (6,3,+\tfrac{1}{3}) =
2 \left( 4 \lambda_2 + 3 \lambda_4 + 16 \lambda'_{4} \right) |\sigma|^2 
+ 2 \left(\omega_{BL}+\omega _R\right) \left(2 \beta'_4 \left(2 \omega_{BL}+\omega _R\right)-\tau \right) \, , \nn \\
& M^2 (\overline{6},1,-\tfrac{4}{3}) = 
2 \left( 4 \lambda_2 + 3 \lambda_4 + 16 \lambda'_{4} \right) |\sigma|^2 
+ 4 \omega_{BL} \left(2 \beta'_4 \left(\omega_{BL}+2 \omega _R\right)-\tau \right)
 \, , \nn \\
& M^2 (\overline{6},1,-\tfrac{1}{3}) = 
4 \left( 3 \lambda_2  + 3 \lambda_4 + 4 \lambda'_{4} \right) |\sigma|^2
- 2 \tau  \left(2 \omega_{BL}+\omega _R\right)
+ 2 \beta _4 \omega _R^2 
+ 4 \beta'_4 \left(3 \omega _R \omega_{BL}+2 \omega_{BL}^2+\omega _R^2\right)
 \, , \nn \\
& M^2 (\overline{6},1,+\tfrac{2}{3}) =
8 \left( 2 \lambda_2 + 3 \lambda_4 + 2 \lambda'_{4} \right) |\sigma|^2 
- 4 \left(\omega_{BL}+\omega _R\right) \left(\tau -2
   \beta'_4 \omega_{BL}\right) \, , \nn
\end{align}
\end{widetext}
while those developing higher-dimensional mass matrices are:
\begin{widetext}
\begin{multline}
M^2 (1,2,+\tfrac{1}{2}) = \\
\left(
\begin{array}{c}
4 (3 \lambda_2 +3 \lambda_4+4 \lambda'_4) |\sigma|^2 
-3 \tau  \left(\omega_{BL}+\omega _R\right) 
+ \frac{\beta _4}{2} \left(-4 \omega _R \omega_{BL}+7\omega_{BL}^2+\omega _R^2\right) + 3 \beta'_4 \left(4 \omega _R \omega_{BL}+3 \omega_{BL}^2+\omega _R^2\right) \\ 
2 \gamma^*_2 \left(\omega _R^2-\omega_{BL}^2\right)
\end{array}
\right. \\ 
\left.
\begin{array}{c}
2 \gamma _2 \left(\omega _R^2-\omega_{BL}^2\right) \\ 
2 (4 \lambda_2+3 \lambda_4 -8 \lambda'_4) |\sigma|^2 - \tau  \left(3 \omega_{BL}+\omega _R\right)
+ \frac{\beta _4}{2} \left(4 \omega _R \omega_{BL}+7 \omega_{BL}^2+\omega _R^2 \right) + 3 \beta'_4 \left(4 \omega _R \omega_{BL}+3 \omega_{BL}^2+\omega _R^2\right)
\end{array}
\right)
\label{matrix:12p12}
\, , 
\end{multline}
\begin{multline}
M^2 (3,2,+\tfrac{7}{6}) = 
\left(
\begin{array}{c}
2 \left( 4 \lambda_2 + 3 \lambda_4 + 16 \lambda'_{4} \right) |\sigma|^2
- \tau  \left(\omega_{BL}+3 \omega _R\right)
+\tfrac{\beta _4}{2} \left(\omega_{BL}-\omega_R\right){}^2
+ \beta'_4 \left(16 \omega _R \omega_{BL}+5 \omega_{BL}^2+3 \omega _R^2\right)
\\ 
2 \gamma^*_2 \left(\omega _R^2-\omega_{BL}^2\right)
\end{array}
\right. \\
\left.
\begin{array}{c}
2 \gamma_2 \left(\omega _R^2-\omega_{BL}^2\right)
\\ 
4 \left( 3 \lambda_2  + 3 \lambda_4 + 4 \lambda'_{4} \right) |\sigma|^2 
- \tau  \left(5 \omega_{BL}+\omega _R\right)
+\tfrac{\beta _4}{2} \left(\omega_{BL}+\omega_R\right){}^2
+ \beta'_4 \left(16 \omega _R \omega_{BL}+5 \omega_{BL}^2+3 \omega _R^2\right)
\end{array}
\right)
\, , 
\end{multline}
\begin{multline}
\label{M82p12}
M^2 (8,2,+\tfrac{1}{2}) =   
\left(
\begin{array}{c}
4 \left( 3 \lambda_2  + 3 \lambda_4 + 4 \lambda'_{4} \right) |\sigma|^2
+ \frac{1}{2} \left(\omega_{BL}+\omega _R\right) \left(6 \beta'_4 \left(3 \omega_{BL}+\omega _R\right)
+\beta _4 \left(\omega_{BL}+\omega_R\right)-6 \tau \right)
\\ 
2 \gamma^*_2 \left(\omega _R^2-\omega_{BL}^2\right)
\end{array}
\right. \\
\left.
\begin{array}{c}
2 \gamma_2 \left(\omega _R^2-\omega_{BL}^2\right)
\\ 
2 \left( 4 \lambda_2 + 3 \lambda_4 + 16 \lambda'_{4} \right) |\sigma|^2
- \tau  \left(3 \omega_{BL}+\omega _R\right)
+\tfrac{\beta _4}{2} \left(\omega_{BL}-\omega_R\right){}^2
+ 3 \beta'_4 \left(4 \omega _R \omega_{BL}+3 \omega_{BL}^2+\omega _R^2\right)
\end{array}
\right)
\, , 
\end{multline}
\begin{multline}
M^2 (\overline{3},1,+\tfrac{1}{3}) =   
\left(
\begin{array}{c}
2  (4 \lambda_2+3 \lambda_4+8 \lambda'_4) |\sigma|^2
-2 \tau \left(\omega_{BL}+\omega _R\right)
+\beta _4 \left(\omega_{BL}^2+\omega _R^2\right)
+ 4 \beta'_4 \left(3 \omega _R \omega_{BL}+2 \omega_{BL}^2+\omega _R^2\right)
\\ 
4 \gamma^*_2 \left(\omega_{BL}^2-\omega _R^2\right) 
\\ 
2 \sqrt{2} \left(-8 \lambda'_4 |\sigma|^2 + \beta _4 \omega _R \omega_{BL} \right)
\end{array}
\right. \\
\left.
\begin{array}{c}
4 \gamma_2 \left(\omega_{BL}^2-\omega _R^2\right) 
\\
4 \left( 3 \lambda_2  + 3 \lambda_4 + 4 \lambda'_{4} \right) |\sigma|^2
-2 \tau \left(2 \omega_{BL}+\omega _R\right) 
+\beta _4 \left(\omega_{BL}^2+\omega _R^2\right)
+4 \beta'_4 \left(3 \omega _R \omega_{BL}+2 \omega_{BL}^2+\omega _R^2\right)
\\
0
\end{array}
\right. \\
\left.
\begin{array}{c}
2 \sqrt{2} \left(-8 \lambda'_4 |\sigma|^2 + \beta _4 \omega _R \omega_{BL} \right)
\\ 
0
\\ 
2 (4 \lambda_2 + 3 \lambda_4) |\sigma|^2
+ 2 \left(
2 \beta'_4 \left(3 \omega _R \omega_{BL}+2 \omega_{BL}^2+\omega _R^2\right)+\beta _4 \left(\omega_{BL}^2+\omega _R^2\right)-\tau\left(\omega_{BL}+\omega _R\right)
\right)
\end{array}
\right)
\, , 
\end{multline}
\end{widetext}
where the mass matrices above are spanned on the following bases (listing only the column basis vectors; the rows are just their  conjugates): $\left\{ (1,2,-\tfrac{1}{2})_{\Sigma^*}, (1,2,-\tfrac{1}{2})_{\Sigma} \right\}$, $\left\{ (\overline{3},2,-\tfrac{7}{6})_{\Sigma^*}, (\overline{3},2,-\tfrac{7}{6})_{\Sigma} \right\}$, $\left\{ (8,2,-\tfrac{1}{2})_{\Sigma^*}, (8,2,-\tfrac{1}{2})_{\Sigma} \right\}$ and $\left\{ (3,1,-\tfrac{1}{3})^{1}_{\Sigma^*}, (3,1,-\tfrac{1}{3})^{1}_{\Sigma}, (3,1,-\tfrac{1}{3})^{3}_{\Sigma^*} \right\}$. 
For potentially ambiguous cases we use superscripts to indicate the $SU(2)_R$ origin of the relevant components.

\subsubsection{Mixed states with components in both $45_H$ and $126_H$}
Finally, the remaining components of  $126_H$ that mix with those in $45_H$  are listed below.
\begin{widetext}
\bea
&& M^2 (1,1,+1) =  
\left(
\begin{array}{cc}
2 \left( - a_2 \omega_{BL} (\omega_{BL} + \omega_R) + \left( \beta_4 - 2 \beta'_4 \right) |\sigma|^2 \right) 
&  2 \left(2 \beta'_4 \left(3 \omega_{BL}+\omega _R\right)+\beta _4 \omega _R-\tau \right) \sigma^* \\
2 \left(2 \beta'_4 \left(3 \omega_{BL}+\omega _R\right)+\beta _4 \omega _R-\tau \right) \sigma 
& 2 \omega _R \left(2 \beta'_4 \left(3 \omega_{BL}+\omega _R\right)+\beta _4 \omega _R-\tau \right)
\end{array}
\right) \, , 
\eea
\bea
&& M^2 (\overline{3},1,-\tfrac{2}{3}) =  
\left(
\begin{array}{cc}
2 \left( - a_2 \omega_R (\omega_{BL} + \omega_R) + \left( \beta_4 - 2 \beta'_4 \right) |\sigma|^2 \right)
& 
-2 \left(4 \beta'_4 \left(\omega_{BL}+\omega _R\right)+\beta _4 \omega_{BL}-\tau \right) \sigma
\\
-2 \left(4 \beta'_4 \left(\omega_{BL}+\omega _R\right)+\beta _4 \omega_{BL}-\tau \right) \sigma^*
& 
2 \omega_{BL} \left(4 \beta'_4 \left(\omega_{BL}+\omega _R\right)+\beta _4 \omega_{BL}-\tau \right)
\end{array}
\right) \, , 
\eea
\begin{multline}
M^2 (3,2,+\tfrac{1}{6}) = \\  
\left(
\begin{array}{cc}
2 \left( - 2 a_2 \omega_{BL} \omega_R + \left( \beta_4 - 2 \beta'_4 \right) |\sigma|^2 \right)
&
\left(2 \beta'_4 \left(5 \omega_{BL}+3 \omega _R\right)+\beta _4 \left(\omega_{BL}+\omega _R\right)-2 \tau \right) \sigma^*
\\
\left(2 \beta'_4 \left(5 \omega_{BL}+3 \omega _R\right)+\beta _4 \left(\omega_{BL}+\omega _R\right)-2 \tau \right) \sigma
&
2 \left(\omega_{BL}+\omega _R\right) \left( 2 \beta'_4 \left(5 \omega_{BL}+3 \omega _R\right)+\beta _4 \left(\omega_{BL}+\omega _R\right)-2 \tau \right)
\\
4 \gamma_2^*  \left(\omega _R-\omega_{BL}\right) \sigma 
&
2 \gamma^* _2 \left(\omega _R^2-\omega_{BL}^2\right)
\end{array}
\right. \\
\left.
\begin{array}{c}
4 \gamma_2  \left(\omega _R-\omega_{BL}\right) \sigma^* 
\\
2 \gamma_2 \left(\omega _R^2-\omega_{BL}^2\right)
\\ 
8 \left( 2 \lambda_2 + 3 \lambda_4 + 2 \lambda'_{4} \right) |\sigma|^2
- \tau  \left(5 \omega_{BL}+3 \omega _R\right)
+\tfrac{\beta _4}{2} \left(\omega_{BL}-\omega_R\right){}^2
+ \beta'_4 \left(8 \omega _R \omega_{BL}+5 \omega_{BL}^2+3 \omega _R^2\right)
\end{array}
\right)
\, , 
\end{multline}
\begin{multline}
\label{Msing110}
M^2 (1,1,0) = \\
\left(
\begin{array}{cc}
2 \left( 12 a_0 \omega_{BL}^2 +  a_2 (\omega_{BL} - \omega_R) (2 \omega_{BL} + \omega_R) - 6 \beta'_4 |\sigma|^2 \right) 
& 4 \sqrt{6} \left( 2 a_0 \omega_{BL} \omega_R - \beta'_4 |\sigma|^2 \right)   
\\
4 \sqrt{6} \left( 2 a_0 \omega_{BL} \omega_R - \beta'_4 |\sigma|^2 \right) 
& 2 \left( 8 a_0 \omega_R^2 - a_2 (\omega_{BL} - \omega_R) (\omega_{BL} + 2 \omega_R) - 4 \beta'_4 |\sigma|^2 \right) 
\\ 
\sqrt{6} \left(\tau +2 \alpha  \omega_{BL} -4 \beta'_4 \left(\omega_{BL}+\omega _R\right) \right) \sigma^*
& 2 \left(\tau +2 \alpha  \omega _R -2 \beta'_4 \left(3 \omega_{BL}+\omega _R\right) \right) \sigma^*
\\
\sqrt{6} \left(\tau +2 \alpha  \omega_{BL} -4 \beta'_4 \left(\omega_{BL}+\omega _R\right) \right) \sigma
& 2 \left(\tau +2 \alpha  \omega _R -2 \beta'_4 \left(3 \omega_{BL}+\omega _R\right) \right) \sigma
\end{array} 
\right. \\
\left. 
\begin{array}{cc}
\sqrt{6} \left(\tau +2 \alpha  \omega_{BL} -4 \beta'_4 \left(\omega_{BL}+\omega _R\right) \right) \sigma 
& \sqrt{6} \left(\tau +2 \alpha  \omega_{BL} -4 \beta'_4 \left(\omega_{BL}+\omega _R\right) \right) \sigma^* \\
2 \left(\tau +2 \alpha  \omega _R -2 \beta'_4 \left(3 \omega_{BL}+\omega _R\right) \right) \sigma 
& 2 \left(\tau +2 \alpha  \omega _R -2 \beta'_4 \left(3 \omega_{BL}+\omega _R\right) \right) \sigma^* \\
4 \lambda_0 |\sigma|^2 & 4 \lambda_0 \sigma^{*2} \\
4 \lambda_0 \sigma^2 & 4 \lambda_0 |\sigma|^2 
\end{array}
\right)
\, ,
\end{multline}
\end{widetext}
Here the relevant bases  are: $\left\{ (\overline{3},1,-\tfrac{2}{3})_{\phi}, (\overline{3},1,-\tfrac{2}{3})_{\Sigma^*} \right\}$ and $\left\{ (\overline{3},2,-\tfrac{1}{6})_{\phi}, (\overline{3},2,-\tfrac{1}{6})_{\Sigma^*}, (\overline{3},2,-\tfrac{1}{6})_{\Sigma} \right\}$ for the coloured triplets and  $\left\{ (1,1,-1)_{\phi}, (1,1,-1)_{\Sigma^*} \right\}$ and $\left\{ (1,1,0)^{15}_{\phi}, (1,1,0)^{1}_{\phi}, (1,1,0)_{\Sigma^*}, (1,1,0)_{\Sigma} \right\}$ for the singlets.
Wherever ambiguous, the superscripts denote the $SU(4)_C$ origin of the relevant components.

Finally, implementing the remaining stationary condition in~\eq{stateq45_2}
(substituting for $a_2$) one obtains
\bea
&& \text{Rank}\,  M^2(1,1,+1) = 1 \, , \nn \\
&& \text{Rank}\,  M^2(3,1,+\tfrac{2}{3}) = 1 \, , \\
&& \text{Rank}\, M^2(3,2,+\tfrac{1}{6}) = 2 \, , \nn \\
&& \text{Rank}\, M^2(1,1,0) = 3 \nn \, . 
\eea
Together with $M^2(3,2,-\tfrac{5}{6}) = 0$, cf.~\eq{Goldstones1}, we account for exactly 
33 WBG bosons corresponding to the coset $SO(10)/SU(3)_c \otimes SU(2)_L \otimes U(1)_Y$.

\subsection{Basic consistency checks}
In order to crosscheck the results given above, we shall now study the scalar spectrum of~\app{scalspectSM} 
in three physically interesting limits. 
In each case one should observe a proper re-clustering of the SM multiplets according to the enhanced 
symmetry as well as extra WGBs.

\subsubsection{The flipped $5' \, 1_{Z'}$ limit, $\omega_R = - \omega_{BL}\neq 0$ and $\sigma=0$}
\label{omegaR=omageBL}
Labelling the scalar states according to the flipped $5' \, 1_{Z'}$ algebra, cf. TABLE~\ref{tab:45decomp}, the $45_H$ components cluster as follows
\begin{align}
& M^2 (24,0) = 4 a_2 \omega^2 \, , \nn \\
& M^2 (10,-4) = 0  \, , \\
& M^2 (1,0) = 4 (10 a_0 + a_2) \omega^2 \, . \nn 
\end{align}
Similarly, for the $126_H$ components we get:
\begin{align}
& M^2(1,+10) = -\nu ^2 +5  \left( \left( \alpha  -4 \beta'_4 \right) \omega -\tau \right) \omega \, , \nn \\
& M^2(\overline{5},+2) = -\nu ^2 + \left( \left(5 \alpha +6 \beta _4 +4 \beta'_4\right) \omega -\tau \right)  \omega  \, , \nn \\
& M^2(10,+6) = -\nu ^2 +  \left( \left(5 \alpha +2 \beta _4 -4 \beta'_4 \right) \omega -3 \tau \right) \omega  \, , \nn \\
& M^2(\overline{15},-6) = -\nu ^2 +  \left( \left( 5 \alpha  -4  \beta'_4 \right)  \omega +3 \tau \right) \omega  \, ,  \quad \\
& M^2(45,-2) = -\nu ^2 + \left(  \left(5 \alpha +2 \beta _4 +4 \beta'_4\right) \omega +\tau \right) \omega  \, , \nn \\
& M^2(\overline{50},+2) = -\nu ^2 +  \left( \left( 5 \alpha  +4 \beta'_4 \right)  \omega -\tau \right) \omega  \nn \, .
\end{align}
As
expected, there are $45-25=20$ WGBs. 

\subsubsection{The $3_c2_L2_R1_{BL}$ limit, $\omega_{BL}\neq 0$, $\omega_R=0$ and $\sigma=0$}
\label{omegaRandsigma=0}
Labelling the scalar states according to the $3_c2_L2_R1_{BL}$ algebra, the $45_H$ components cluster as follows:
\begin{align}
&M^2(1,3,1,0)= - 2 a_2 \omega_{BL}^2 \, , \nn \\
&M^2(1,1,3,0)= - 2 a_2 \omega_{BL}^2 \, , \nn \\
&M^2(8,1,1,0)= 4 a_2 \omega_{BL}^2 \, , \\
&M^2(3,2,2,-\tfrac{1}{3})= 0 \, , \nn \\
&M^2(\overline{3},1,1,-\tfrac{2}{3})= 0 \, , \nn \\
&M^2(1,1,1,0)= 4 \left(6 a_0+a_2\right) \omega_{BL}^2 \, . \nn
\end{align}
Analogously, for the $126_H$ components we get
\begin{align}
&\!\!\!\!\!\!\!\! M^2(1,3,1,-1) = 
-\nu ^2 
+ 3  \left( \left( \alpha  - 2 \beta'_4 \right) \omega_{BL} - \tau \right) \omega_{BL} \, , \nn \\
&\!\!\!\!\!\!\!\! M^2(1,1,3,+1) = 
-\nu ^2
+ 3 \left( \left(\alpha -2 \beta'_4\right) \omega_{BL} + \tau \right) \omega_{BL} 
\, , \nn \\
&\!\!\!\!\!\!\!\! M^2(3,3,1,-\tfrac{1}{3}) = 
-\nu ^2 \nn \\
&\!\!\!\!\!\!\!\! \qquad\qquad\qquad\ \; + \left( \left(3 \alpha  +2 \left(\beta _4 + \beta'_4\right) \right) \omega_{BL} -\tau \right) \omega_{BL} 
\, , \nn \\
&\!\!\!\!\!\!\!\! M^2(\overline{3},1,3,+\tfrac{1}{3}) = 
-\nu ^2  \\
&\!\!\!\!\!\!\!\! \qquad\qquad\qquad\ \; + \left( \left(3 \alpha +2 \left(\beta _4 + \beta'_4\right)\right) \omega_{BL}  + \tau \right) \omega_{BL} 
\, , \nn \\
&\!\!\!\!\!\!\!\! M^2(6,3,1,+\tfrac{1}{3}) = 
-\nu ^2
+  \left(  \left( 3 \alpha  +2 \beta'_4 \right) \omega_{BL} + \tau \right) \omega_{BL}
\, , \nn \\
&\!\!\!\!\!\!\!\! M^2(\overline{6},1,3,-\tfrac{1}{3}) = 
-\nu ^2
+ \left( \left( 3 \alpha  +2 \beta'_4 \right) \omega_{BL} -\tau  \right) \omega_{BL}
\, , \nn
\end{align}
\begin{widetext}
\bea
&& \!\!\!\!\!\!\!\!\!\! M^2(1,2,2,0) = 
\left(
\begin{array}{cc}
-\nu ^2 + \frac{1}{2} \left(6 \alpha +7 \beta _4+6 \beta'_4\right) \omega_{BL}^2 & -2 \gamma _2 \omega_{BL}^2 \\
 -2 \gamma^*_2 \omega_{BL}^2 & -\nu ^2 + \frac{1}{2}  \left(6 \alpha +7 \beta _4+6 \beta'_4\right) \omega_{BL}^2
\end{array}
\right)
\, , \\
&& \!\!\!\!\!\!\!\!\!\! M^2(8,2,2,0) = 
\left(
\begin{array}{cc}
-\nu ^2 + \frac{1}{2}  \left(6 \alpha +\beta _4+6 \beta'_4\right) \omega_{BL}^2 & -2 \gamma _2 \omega_{BL}^2 \\
 -2 \gamma^*_2 \omega_{BL}^2 & -\nu ^2 + \frac{1}{2} \left(6 \alpha +\beta _4+6 \beta'_4\right)  \omega_{BL}^2
\end{array}
\right)
\, , \\
&& \!\!\!\!\!\!\!\!\!\! M^2(\overline{3},1,1,+\tfrac{1}{3}) = 
\left(
\begin{array}{cc}
-\nu ^2 + \left( \left(3 \alpha +\beta _4+2 \beta'_4\right) \omega_{BL} + \tau \right) \omega_{BL}  & 4 \gamma _2 \omega_{BL}^2 \\
 4 \gamma^*_2 \omega_{BL}^2 & -\nu ^2 + \left( \left(3 \alpha +\beta _4+2 \beta'_4\right) \omega_{BL} -\tau \right) \omega_{BL}
\end{array}
\right)
\, , \\
&& \!\!\!\!\!\!\!\!\!\! M^2(3,2,2,+\tfrac{2}{3}) = 
\left(
\begin{array}{cc}
-\nu ^2 + \frac{1}{2}  \left( \left(6 \alpha +\beta _4-2 \beta'_4\right) \omega_{BL} + 4 \tau \right) \omega_{BL} & -2 \gamma _2 \omega_{BL}^2 \\
 -2 \gamma^*_2 \omega_{BL}^2 & -\nu ^2 + \frac{1}{2} \left( \left(6 \alpha +\beta _4-2 \beta'_4\right) \omega_{BL} -4 \tau \right) \omega_{BL}
\end{array}
\right)
\, , 
\eea
\end{widetext}
where the   matrices above are spanned over 
$\left\{(3,\!1,\!1,\!-\tfrac{1}{3})_{\Sigma^*}\!,\! (3,\!1,\!1,\!-\tfrac{1}{3})_{\Sigma}\right\}\!,\! 
\left\{(\overline{3},\!2,\!2,\!-\tfrac{2}{3})_{\Sigma^*}\!,\! (\overline{3},\!2,\!2,\!-\tfrac{2}{3})_{\Sigma}\right\}$\!,\!
$\left\{(1,\!2,\!2,\!0)_{\Sigma^*}, (1,\!2,\!2,\!0)_{\Sigma}\right\}$ and
$\left\{(8,\!2,\!2,\!0)_{\Sigma^*}, (8,\!2,\!2,\!0)_{\Sigma}\right\}$. 
As expected, there are $45-15=30$ WGBs in the spectrum.

It is worth noting that $(1,3,1,0)$ and $(1,1,3,0)$ remain degenerate which is due to the fact that
for $\omega_R=0$ the $D$-parity is conserved by even $\omega_{BL}$ powers.
On the contrary, in the $126_H$ components the $D$-parity is broken by the $\tau$ term that is linear in $\omega_{BL}$.
\subsubsection{The $4_C2_L1_R$ limit, $\omega_{R} \neq 0$, $\omega_{BL} = 0$ and $\sigma = 0$}
\label{omegaBLandsigma=0}
Again, as anticipated, the clustering of the scalar spectrum in the $4_C2_L1_R$ limit follows the decomposition rule listed in TABLE~\ref{tab:45decomp}:
\begin{align}
&M^2(15,1,0)= - 2 a_2 \omega _R^2 \, , \nn \\
&M^2(1,3,0)= 4 a_2 \omega _R^2 \, , \nn \\
&M^2(6,2,+\tfrac{1}{2})= 0 \, , \\
&M^2(1,1,+1)= 0 \, , \nn \\
&M^2(1,1,0)= 4 \left(4 a_0+a_2\right) \omega _R^2 \, , \nn
\end{align}
\begin{align}
&M^2(10,3,0)= 
-\nu ^2 + 2 \left(\alpha + \beta'_4\right) \omega _R^2 
\, , \nn \\
&M^2(\overline{10},1,-1)= 
-\nu ^2 + 2 \left( \left(\alpha - \beta'_4\right) \omega _R +\tau \right) \omega _R
\, , \\
&M^2(\overline{10},1,0)= 
-\nu ^2 + 2 \left(\alpha + \beta _4 + \beta'_4\right) \omega _R^2 
\, , \nn \\
&M^2(\overline{10},1,+1)= 
-\nu ^2 + 2 \left( \left(\alpha - \beta'_4 \right) \omega _R - \tau \right) \omega _R
\, , \nn
\end{align}
\begin{widetext}
\bea
&& M^2(6,1,0)=
\left(
\begin{array}{cc}
-\nu ^2 + \left(2 \alpha +\beta _4+2 \beta'_4\right) \omega _R^2  & -4 \gamma _2 \omega _R^2 \\
 -4 \gamma^*_2 \omega _R^2 & -\nu ^2 + \left(2 \alpha +\beta _4+2 \beta'_4\right) \omega _R^2
\end{array}
\right) 
\, , \\
&& M^2(15,2,+\tfrac{1}{2})=
\left(
\begin{array}{cc}
-\nu ^2 + \frac{1}{2}  \left( \left(4 \alpha +\beta _4+2 \beta'_4\right) \omega _R  - 2 \tau \right) \omega _R & 2 \gamma _2 \omega _R^2 \\
 2 \gamma^*_2 \omega _R^2 & -\nu ^2 + \frac{1}{2} \left( \left(4 \alpha +\beta _4+2 \beta'_4\right) \omega _R + 2 \tau \right) \omega _R
\end{array}
\right)
\, . 
\eea
\end{widetext}
The mass matrices above are defined on the bases $\left\{(6,1,0)_{\Sigma^*}, (6,1,0)_{\Sigma}\right\}$ and $\left\{(15,2,-\tfrac{1}{2})_{\Sigma^*}, (15,2,+\tfrac{1}{2})_{\Sigma}\right\}$, respectively.
As expected, there are in total $45-19=26$ WGBs.

\subsection{Few remarks on global symmetries}
\label{globalsymmetries}
It is illuminating to look at the global symmetries of the scalar potential when only the moduli of $45_H$ and $126_H$ appear in the scalar potential.
In such a case, 
(i.e., with $a_2=\lambda_2=\lambda_4=\lambda'_4=\eta_2=\tau=\beta_4=\beta'_4=\gamma_2=0$),
the global symmetry of $V_{0}$ in~\eq{scalpotgen}
is $O(45)\otimes O(252)$.
This symmetry is spontaneously broken into $O(44)\otimes O(251)$ by the $45_H$ and $126_H$ VEVs
yielding $44+251=295$ Goldstones in the scalar spectrum.
Since, at the same time, the gauge $SO(10)$ symmetry is broken to the SM gauge group,
$45-12=33$ would-be Goldstone bosons
are ``eaten'' by the gauge fields associated to the $SO(10)/\text{SM}$ coset and drop from  the scalar spectrum,
$295-33=262$ PGB remain.
Their masses are generally expected to receive contributions
from the explicitly breaking terms 
$a_2$, $\lambda_2$, $\lambda_4$, $\lambda'_4$, $\eta_2$, $\tau$, $\beta_4$, $\beta'_4$ and $\gamma_2$.
A detailed inspection of the mass matrices in~\eqs{M2130}{Msing110}, indeed, reveals the total of 262 massless degrees of freedom. 

In this respect, let us emphasize that the $(1,3,0)$ and $(8,1,0)$ components of $45_{H}$ of our central interest in Section~\ref{treescalarspectrum} belong to this category but, for various reasons, they receive just the $a_{2}$-proportional mass contribution at the tree level while the other couplings enter the relevant mass formulae only via loops.

\section{The tree-level gauge boson spectrum}
\label{gaugespectrum}
Let us start with the scalar kinetic 
terms\footnote{Notice that $\Sigma^*_{abcde} \Sigma_{abcde} = \phi_{abcde} \phi_{abcde}$, 
where $\Sigma$ and $\Sigma^*$ are defined respectively in~\eqs{defsigma}{defsigmabar}.} 
\be
\label{kinetic45}
\frac{1}{4} (D_{\mu}\phi)^*_{ab} (D^{\mu}\phi)_{ab}\;\;\;\;
\text{and}\;\;\;\;
\frac{1}{2 \cdot 5!}(D_{\mu}\Sigma)^*_{abcde} (D^{\mu}\Sigma)_{abcde} \,,
\ee
where the covariant derivatives are given by
\be
(D_{\mu}\phi)_{ab} = 
\partial_{\mu}\phi_{ab} -i
\frac{1}{2}g(A_\mu)_{ij} \left[ \epsilon_{ij}, \phi \right]_{ab}
\ee
and
\begin{align}
(D_{\mu} \Sigma)_{abcde} & = 
\partial_{\mu}\Sigma_{abcde}  
-i \frac{1}{2}g(A_\mu)_{ij} 
\Bigl[ (\epsilon_{ij})_{aa'} \Sigma_{a'bcde} \nn \\
& + (\epsilon_{ij})_{bb'} \Sigma_{ab'cde} + (\epsilon_{ij})_{cc'} \Sigma_{abc'de}\\ 
& + (\epsilon_{ij})_{dd'} \Sigma_{abcd'e} + (\epsilon_{ij})_{ee'} \Sigma_{abcde'} \Bigr] \, ,\nn
\end{align}
respectively, and $\epsilon_{ij}$ ($i,j = 1, \ldots ,10$) are the $SO(10)$ generators in the 
fundamental representation 
\be
\label{SO10gen10}
(\epsilon_{ij})_{ab}=-i(\delta_{ai}\delta_{bj} - \delta_{aj}\delta_{bi}) \, .
\ee 
One finds the following expressions for the field dependent mass matrices of the gauge bosons
\be
\label{fielddepmass45}
\mathcal{M}_A^2(\phi)_{(ij)(kl)}
= \frac{g^2}{2} \Tr \left[ \epsilon_{(ij)}, \phi \right] \left[ \epsilon_{(kl)}, \phi \right]
\ee
for the contribution from the VEVs in the $45_{H}$, 
and 
\begin{multline}
\label{fielddepmass126}
\mathcal{M}_A^2(\Sigma, \Sigma^*)_{(ij)(kl)} = 
- \frac{g^2}{2 \cdot 5!} 
\Bigl[ (\epsilon_{(ij)})_{aa'} \Sigma^*_{a'bcde} \nn\\
+ (\epsilon_{(ij)})_{bb'} \Sigma^*_{ab'cde} + (\epsilon_{(ij)})_{cc'} \Sigma^*_{abc'de} \nn\\
+ (\epsilon_{(ij)})_{dd'} \Sigma^*_{abcd'e} + (\epsilon_{(ij)})_{ee'} \Sigma^*_{abcde'} \Bigr] \\
\times \Bigl[ (\epsilon_{(kl)})_{aa''} \Sigma_{a''bcde}\nn\\
 + (\epsilon_{(kl)})_{bb''} \Sigma_{ab''cde} + (\epsilon_{(kl)})_{cc''} \Sigma_{abc''de} \nn\\
+ (\epsilon_{(kl)})_{dd''} \Sigma_{abcd''e} + (\epsilon_{(kl)})_{ee''} \Sigma_{abcde''} \Bigr] + \text{c.c.} 
\, ,
\end{multline}
for that of the VEV in $126_{H}$,
where $(ij)$, $(kl)$ stand for ordered pairs of indices.

\subsection{Contributions to gauge bosons masses from $45_H$}
\label{gauge45}
Evaluating~\eq{fielddepmass45} one obtains
\begin{align}
&\mathcal{M}_A^2(1,1,+1)=4 g^2 \omega _R^2 \, ,
\nn \\[0ex]
&\mathcal{M}_A^2(\overline{3},1,-\tfrac{2}{3})=4 g^2 \omega_{BL}^2 \, ,
\nn\\[0ex]
&\mathcal{M}_A^2(3,2,-\tfrac{5}{6})=g^2 \left(\omega _R-\omega_{BL}\right)^2 \, ,
\nn\\[0ex]
&\mathcal{M}_A^2(3,2,+\tfrac{1}{6})=g^2 \left(\omega _R+\omega_{BL}\right)^2 \, ,
\end{align}
while, as expected, there are no contributions of $\vev{45_{H}}$ to $\mathcal{M}_A^2(1,3,0)$, $\mathcal{M}_A^2(8,1,0)$ and $\mathcal{M}_A^2(1,1,0)$.
Note that, in the limits of the standard $5 \, 1_{Z}$ $(\omega_R = \omega_{BL})$, flipped $5' \, 1_{Z'}$ $(\omega_R = -\omega_{BL})$,
$3_c\, 2_L\, 2_R\, 1_{BL}$ ($\omega_R=0$) and $4_C\, 2_L\, 1_R$ ($\omega_{BL}=0$) vacua,  there are 
25, 25, 15 and 19 massless gauge bosons, respectively.

\subsection{Contributions to gauge bosons masses from $126_H$}
\label{gauge126}
Besides $\mathcal{M}_A^2(1,3,0)$, $\mathcal{M}_A^2(8,1,0)$ and $\mathcal{M}_A^2(3,2,-\tfrac{5}{6})$ which receive no contributions from  $\vev{126_H}$ one has  
\begin{align}
\label{gaugespectrum16}
&\mathcal{M}_A^2(1,1,+1)= 2 g^2 |\sigma|^2 \, ,
 \nn\\[0ex]
&\mathcal{M}_A^2(\overline{3},1,-\tfrac{2}{3})= 2 g^2 |\sigma|^2 \, ,
\nn\\[0ex]
&\mathcal{M}_A^2(3,2,+\tfrac{1}{6})= 2 g^2 |\sigma|^2 \, ,
\nn\\[0ex]
&\mathcal{M}_A^2(1,1,0)= 
\left(
\begin{array}{cc}
 6  & 2 \sqrt{6}  \\
 2 \sqrt{6}  & 4
\end{array}
\right)g^2 |\sigma|^2 \,.
\nn
\end{align}
Here the SM singlet matrix is defined on the pair of singlets from $15$ and $1$ of $SU(4)_{C}$, respectively. One has
\begin{align}
&\text{Det} \mathcal{M}_A^2(1,1,0)=0 \, , \\[0ex]
&\Tr \mathcal{M}_A^2(1,1,0)=10 g^2 |\sigma|^2 \, ,
\end{align}
and, as expected, there is a massless state in the singlet sector.
The number of vanishing entries corresponds to the dimension
of the $SU(5)$ algebra
preserved by the $126_H$ VEV $\sigma$.
Summing together the $45_H$ and $126_H$ contributions, one ends up with 12 massless vector bosons of the unbroken $SU(3)_{c}\otimes SU(2)_{L}\otimes U(1)_{Y}$ gauge symmetry.

\section{Sample spectra}
\label{samplespectrum}
The details of the bosonic spectrum for the sample settings described in Sects.~\ref{sect:specific63p13} and~\ref{sect:specific82p12} are given in TABLEs~\ref{sample63p13} and~\ref{sample82p12}, respectively.  The symbols RS, CS and VB denote, consecutively, real scalars, complex scalars and vector bosons. Notice that a double counting in the $SU(2)_{L}$ doublet sector is, conveniently, avoided by treating both $(1,2,+\tfrac{1}{2})$'s as real scalars; for further details see Sect.~\ref{sect:technicalities}. 

Note also that since we are working in the Feynman gauge of the $SO(10) \to \text{SM}$ broken phase, the Goldstones affect the $b$-coefficients at the mass scales of the corresponding vector bosons (recall the Feynman-gauge pole structure of the Goldstone boson propagators); hence, the Goldstone bosons (GB) are displayed along with the relevant gauge fields in TABLEs~\ref{sample63p13} and~\ref{sample82p12}. 
\renewcommand{\arraystretch}{1.0}
\begin{table}[t]
\begin{tabular}{c|c|c|c|c}
\hline\hline
multiplet & \; type\;  & eigenstate& $\Delta b^{321}$ & mass [GeV]  \\
\hline
$\bf (6,3,+\tfrac{1}{3})$ & CS & 1 & $(\tfrac{5}{2}, 4, \tfrac{2}{5})$ & $\bf 5.6 \times 10^{11}$ \\
\hline
$(1,1,-1)$ & VB & 1 & $(0, 0, -\tfrac{11}{5})$ & $1.3 \times 10^{14}$ \\
\hline
$(1,1,+1)$ & VB & 1 & $(0, 0, -\tfrac{11}{5})$ & $1.3 \times 10^{14}$ \\
\hline
$(1,1,+1)$ & GB & 1 & $(0, 0, \tfrac{1}{5})$ & $1.3 \times 10^{14}$ \\
\hline
$(1,1,0)$ & VB & 1 & $(0, 0, 0)$ & $2.8 \times 10^{14}$ \\
\hline
$(1,1,0)$ & GB & 1 & $(0, 0, 0)$ & $2.8 \times 10^{14}$ \\
\hline
$(8,1,0)$ & RS & 1 & $(\tfrac{1}{2}, 0, 0)$ & $7.7 \times 10^{14}$ \\
\hline
$(3,2,+\tfrac{1}{6})$ & CS & 2 & $(\tfrac{1}{3}, \tfrac{1}{2}, \tfrac{1}{30})$ & $1.1 \times 10^{15}$ \\
\hline
$(3,2,+\tfrac{7}{6})$ & CS & 1 & $(\tfrac{1}{3}, \tfrac{1}{2}, \tfrac{49}{30})$ & $1.2 \times 10^{15}$ \\
\hline
$(1,1,0)$ & RS & 2 & $(0, 0, 0)$ & $4.3 \times 10^{15}$ \\
\hline
$(1,1,+2)$ & CS & 1 & $(0, 0, \tfrac{4}{5})$ & $4.5 \times 10^{15}$ \\
\hline
$\bf(\overline{3},2,-\tfrac{1}{6})$ & \bf VB & 1 & $(-\tfrac{11}{3}, -\tfrac{11}{2}, -\tfrac{11}{30})$ & $\bf5.2 \times 10^{15}$ \\
\hline
$\bf(3,2,+\tfrac{1}{6})$ & \bf VB & 1 & $(-\tfrac{11}{3}, -\tfrac{11}{2}, -\tfrac{11}{30})$ & $\bf5.2 \times 10^{15}$ \\
\hline
$\bf (3,2,+\tfrac{1}{6})$ & \bf GB & 1 & $(\tfrac{1}{3}, \tfrac{1}{2}, \tfrac{1}{30})$ & $\bf 5.2 \times 10^{15}$ \\
\hline
$\bf(\overline{3},2,+\tfrac{5}{6})$ & \bf VB & 1 & $(-\tfrac{11}{3}, -\tfrac{11}{2}, -\tfrac{55}{6})$ & $\bf5.2 \times 10^{15}$ \\
\hline
$\bf(3,2,-\tfrac{5}{6})$ & \bf VB & 1 & $(-\tfrac{11}{3}, -\tfrac{11}{2}, -\tfrac{55}{6})$ & $\bf5.2 \times 10^{15}$ \\
\hline
$\bf (3,2,-\tfrac{5}{6})$ & \bf GB & 1 & $(\tfrac{1}{3}, \tfrac{1}{2}, \tfrac{5}{6})$ & $\bf 5.2 \times 10^{15}$ \\
\hline
$(1,1,+1)$ & CS & 2 & $(0, 0, \tfrac{1}{5})$ & $5.6 \times 10^{15}$ \\
\hline
$(1,1,0)$ & RS & 3 & $(0, 0, 0)$ & $5.7 \times 10^{15}$ \\
\hline
$(1,3,0)$ & RS & 1 & $(0, \tfrac{1}{3}, 0)$ & $6.1 \times 10^{15}$ \\
\hline
$(\overline{3},1,+\tfrac{1}{3})$ & CS & 1 & $(\tfrac{1}{6}, 0, \tfrac{1}{15})$ & $6.4 \times 10^{15}$ \\
\hline
$(8,2,+\tfrac{1}{2})$ & CS & 1 & $(2, \tfrac{4}{3}, \tfrac{4}{5})$ & $9.3 \times 10^{15}$ \\
\hline
$(\overline{3},1,+\tfrac{4}{3})$ & CS & 1 & $(\tfrac{1}{6}, 0, \tfrac{16}{15})$ & $9.6 \times 10^{15}$ \\
\hline
$(\overline{3},1,+\tfrac{1}{3})$ & CS & 2 & $(\tfrac{1}{6}, 0, \tfrac{1}{15})$ & $9.6 \times 10^{15}$ \\
\hline
$(\overline{3},1,-\tfrac{2}{3})$ & CS & 2 & $(\tfrac{1}{6}, 0, \tfrac{4}{15})$ & $9.6 \times 10^{15}$ \\
\hline
$(\overline{3},1,-\tfrac{2}{3})$ & VB & 1 & $(-\tfrac{11}{6}, 0, -\tfrac{44}{15})$ & $1.0 \times 10^{16}$ \\
\hline
$(3,1,+\tfrac{2}{3})$ & VB & 1 & $(-\tfrac{11}{6}, 0, -\tfrac{44}{15})$ & $1.0 \times 10^{16}$ \\
\hline
$(\overline{3},1,-\tfrac{2}{3})$ & GB & 1 & $(\tfrac{1}{6}, 0, \tfrac{4}{15})$ & $1.0 \times 10^{16}$ \\
\hline
$(8,2,+\tfrac{1}{2})$ & CS & 2 & $(2, \tfrac{4}{3}, \tfrac{4}{5})$ & $1.1 \times 10^{16}$ \\
\hline
$(\overline{6},1,+\tfrac{2}{3})$ & CS & 1 & $(\tfrac{5}{6}, 0, \tfrac{8}{15})$ & $1.5 \times 10^{16}$ \\
\hline
$(1,2,+\tfrac{1}{2})$ & {RS} & 1 & $(0, \tfrac{1}{12}, \tfrac{1}{20})$ & $1.5 \times 10^{16}$ \\
\hline
$(\overline{6},1,-\tfrac{1}{3})$ & CS & 1 & $(\tfrac{5}{6}, 0, \tfrac{2}{15})$ & $1.5 \times 10^{16}$ \\
\hline
$(\overline{6},1,-\tfrac{4}{3})$ & CS & 1 & $(\tfrac{5}{6}, 0, \tfrac{32}{15})$ & $1.5 \times 10^{16}$ \\
\hline
$(1,2,+\tfrac{1}{2})$ & {RS} & 2 & $(0, \tfrac{1}{12}, \tfrac{1}{20})$ & $1.6 \times 10^{16}$ \\
\hline
$(\overline{3},1,+\tfrac{1}{3})$ & CS & 3 & $(\tfrac{1}{6}, 0, \tfrac{1}{15})$ & $1.7 \times 10^{16}$ \\
\hline
$(3,3,-\tfrac{1}{3})$ & CS & 1 & $(\tfrac{1}{2}, 2, \tfrac{1}{5})$ & $1.8 \times 10^{16}$ \\
\hline
$(3,2,+\tfrac{1}{6})$ & CS & 3 & $(\tfrac{1}{3}, \tfrac{1}{2}, \tfrac{1}{30})$ & $2.1 \times 10^{16}$ \\
\hline
$(3,2,+\tfrac{7}{6})$ & CS & 2 & $(\tfrac{1}{3}, \tfrac{1}{2}, \tfrac{49}{30})$ & $2.1 \times 10^{16}$ \\
\hline
$(1,3,-1)$ & CS & 1 & $(0, \tfrac{2}{3}, \tfrac{3}{5})$ & $2.6 \times 10^{16}$ \\
\hline
$(1,1,0)$ & RS & 4 & $(0, 0, 0)$ & $3.0 \times 10^{16}$ \\
\hline\hline
\end{tabular}
\caption{\label{sample63p13}
A sample spectrum featuring a light $(6,3,+\tfrac{1}{3})$ threshold. 
The relevant scalar potential parameters are given 
in the left column of~\Table{TableSampleParameters}. $\Delta b^{321}$ indicates 
the shift in the one-loop beta-function. The light threshold and the vector bosons defining the GUT scale are in boldface.}
\end{table}

\begin{table}[t]
\begin{tabular}{c|c|c|c|c}
\hline\hline
multiplet & \; type\;  & eigenstate & $\Delta b^{321}$ & mass [GeV]  \\
\hline
$\bf (8,2,+\tfrac{1}{2})$ & CS & 1 & $(2, \tfrac{4}{3}, \tfrac{4}{5})$ & $\bf 2.3 \times 10^{4}$ \\
\hline
$(\overline{3},1,-\tfrac{2}{3})$ & VB & 1 & $(-\tfrac{11}{6}, 0, -\tfrac{44}{15})$ & $2.8 \times 10^{13}$ \\
\hline
$(3,1,+\tfrac{2}{3})$ & VB & 1 & $(-\tfrac{11}{6}, 0, -\tfrac{44}{15})$ & $2.8 \times 10^{13}$ \\
\hline
$(\overline{3},1,-\tfrac{2}{3})$ & GB & 1 & $(\tfrac{1}{6}, 0, \tfrac{4}{15})$ & $2.8 \times 10^{13}$ \\
\hline
$(1,1,0)$ & VB & 1 & $(0, 0, 0)$ & $6.1 \times 10^{13}$ \\
\hline
$(1,1,0)$ & GB & 1 & $(0, 0, 0)$ & $6.1 \times 10^{13}$ \\
\hline
$(3,2,+\tfrac{7}{6})$ & CS & 1 & $(\tfrac{1}{3}, \tfrac{1}{2}, \tfrac{49}{30})$ & $2.6 \times 10^{14}$ \\
\hline
$(3,2,+\tfrac{1}{6})$ & CS & 3 & $(\tfrac{1}{3}, \tfrac{1}{2}, \tfrac{1}{30})$ & $2.8 \times 10^{14}$ \\
\hline
$(1,2,+\tfrac{1}{2})$ & {RS} & 1 & $(0, \tfrac{1}{12}, \tfrac{1}{20})$ & $3.3 \times 10^{14}$ \\
\hline
$(1,1,0)$ & RS & 2 & $(0, 0, 0)$ & $2.2 \times 10^{15}$ \\
\hline
$(\overline{3},1,-\tfrac{2}{3})$ & CS & 2 & $(\tfrac{1}{6}, 0, \tfrac{4}{15})$ & $2.3 \times 10^{15}$ \\
\hline
$(6,3,+\tfrac{1}{3})$ & CS & 1 & $(\tfrac{5}{2}, 4, \tfrac{2}{5})$ & $2.3 \times 10^{15}$ \\
\hline
$(3,3,-\tfrac{1}{3})$ & CS & 1 & $(\tfrac{1}{2}, 2, \tfrac{1}{5})$ & $2.3 \times 10^{15}$ \\
\hline
$(1,3,-1)$ & CS & 1 & $(0, \tfrac{2}{3}, \tfrac{3}{5})$ & $2.3 \times 10^{15}$ \\
\hline
$(\overline{6},1,-\tfrac{4}{3})$ & CS & 1 & $(\tfrac{5}{6}, 0, \tfrac{32}{15})$ & $3.2 \times 10^{15}$ \\
\hline
$(1,1,0)$ & RS & 3 & $(0, 0, 0)$ & $3.3 \times 10^{15}$ \\
\hline
$(8,1,0)$ & RS & 1 & $(\tfrac{1}{2}, 0, 0)$ & $4.6 \times 10^{15}$ \\
\hline
$(1,3,0)$ & RS & 1 & $(0, \tfrac{1}{3}, 0)$ & $6.1 \times 10^{15}$ \\
\hline
$\bf(\overline{3},2,+\tfrac{5}{6})$ & \bf VB & 1 & $(-\tfrac{11}{3}, -\tfrac{11}{2}, -\tfrac{55}{6})$ & $\bf8.7 \times 10^{15}$ \\
\hline
$\bf(3,2,-\tfrac{5}{6})$ & \bf VB & 1 & $(-\tfrac{11}{3}, -\tfrac{11}{2}, -\tfrac{55}{6})$ & $\bf8.7 \times 10^{15}$ \\
\hline
$\bf (3,2,-\tfrac{5}{6})$ & \bf GB & 1 & $(\tfrac{1}{3}, \tfrac{1}{2}, \tfrac{5}{6})$ & $\bf 8.7 \times 10^{15}$ \\
\hline
$\bf(\overline{3},2,-\tfrac{1}{6})$ & \bf VB & 1 & $(-\tfrac{11}{3}, -\tfrac{11}{2}, -\tfrac{11}{30})$ & $\bf8.7 \times 10^{15}$ \\
\hline
$\bf(3,2,+\tfrac{1}{6})$ & \bf VB & 1 & $(-\tfrac{11}{3}, -\tfrac{11}{2}, -\tfrac{11}{30})$ & $\bf8.7 \times 10^{15}$ \\
\hline
$\bf (3,2,+\tfrac{1}{6})$ & \bf GB & 1 & $(\tfrac{1}{3}, \tfrac{1}{2}, \tfrac{1}{30})$ & $\bf 8.7 \times 10^{15}$ \\
\hline
$(\overline{3},1,+\tfrac{1}{3})$ & CS & 1 & $(\tfrac{1}{6}, 0, \tfrac{1}{15})$ & $1.1 \times 10^{16}$ \\
\hline
$(\overline{3},1,+\tfrac{1}{3})$ & CS & 2 & $(\tfrac{1}{6}, 0, \tfrac{1}{15})$ & $1.2 \times 10^{16}$ \\
\hline
$(1,1,+1)$ & CS & 2 & $(0, 0, \tfrac{1}{5})$ & $1.6 \times 10^{16}$ \\
\hline
$(\overline{3},1,+\tfrac{1}{3})$ & CS & 3 & $(\tfrac{1}{6}, 0, \tfrac{1}{15})$ & $1.6 \times 10^{16}$ \\
\hline
$(\overline{6},1,-\tfrac{1}{3})$ & CS & 1 & $(\tfrac{5}{6}, 0, \tfrac{2}{15})$ & $1.6 \times 10^{16}$ \\
\hline
$(3,2,+\tfrac{7}{6})$ & CS & 2 & $(\tfrac{1}{3}, \tfrac{1}{2}, \tfrac{49}{30})$ & $1.7 \times 10^{16}$ \\
\hline
$(1,2,+\tfrac{1}{2})$ &{RS} & 2 & $(0, \tfrac{1}{12}, \tfrac{1}{20})$ & $1.7 \times 10^{16}$ \\
\hline
$(8,2,+\tfrac{1}{2})$ & CS & 2 & $(2, \tfrac{4}{3}, \tfrac{4}{5})$ & $1.7 \times 10^{16}$ \\
\hline
$(3,2,+\tfrac{1}{6})$ & CS & 2 & $(\tfrac{1}{3}, \tfrac{1}{2}, \tfrac{1}{30})$ & $1.7 \times 10^{16}$ \\
\hline
$(1,1,-1)$ & VB & 1 & $(0, 0, -\tfrac{11}{5})$ & $1.7 \times 10^{16}$ \\
\hline
$(1,1,+1)$ & VB & 1 & $(0, 0, -\tfrac{11}{5})$ & $1.7 \times 10^{16}$ \\
\hline
$(1,1,+1)$ & GB & 1 & $(0, 0, \tfrac{1}{5})$ & $1.7 \times 10^{16}$ \\
\hline
$(1,1,+2)$ & CS & 1 & $(0, 0, \tfrac{4}{5})$ & $2.4 \times 10^{16}$ \\
\hline
$(\overline{3},1,+\tfrac{4}{3})$ & CS & 1 & $(\tfrac{1}{6}, 0, \tfrac{16}{15})$ & $2.4 \times 10^{16}$ \\
\hline
$(\overline{6},1,+\tfrac{2}{3})$ & CS & 1 & $(\tfrac{5}{6}, 0, \tfrac{8}{15})$ & $2.4 \times 10^{16}$ \\
\hline
$(1,1,0)$ & RS & 4 & $(0, 0, 0)$ & $4.1 \times 10^{16}$ \\
\hline\hline
\end{tabular}
\caption{\label{sample82p12}
The same as in~\Table{sample63p13} but for the case discussed in Sect.~\ref{sect:specific82p12} featuring a  light $(8,2,+\tfrac{1}{2})$ threshold. 
The relevant scalar potential parameters are given 
in the right column of~\Table{TableSampleParameters}. 
Notice that, as required by consistency,  in both cases $b_{SM}+\sum \Delta b^{321}=(-\tfrac{37}{3},-\tfrac{37}{3},-\tfrac{37}{3})$.}
\end{table}

\end{document}